\documentclass[12pt]{article}
\usepackage{amsfonts,amssymb,amsmath,fullpage,pst-node,rotating}
\usepackage{multirow}
\newcommand{\R}{{\mathbb R}}
\newcommand{\Z}{{\mathbb Z}}

\newcommand{\mH}{{\cal Q}}

\newcommand{\T}{{\mathbb T}}
\newcommand{\I}{{\mathbb I}}
\newcommand{\V}{{\mathbb V}}
\newcommand{\mO}{{\mathbb O}}
\newcommand{\D}{{\mathbb D}}

\newcommand{\Fix}{{\rm Fix}\,}

\newcommand{\bq}{\bf q}

\newcommand{\rl}{{\,|\,}}

\newtheorem{definition}{Definition}
\newtheorem{theorem}{Theorem}
\newtheorem{lemma}{Lemma}

\newtheorem{remark}{Remark}
\newcommand{\proof}{\noindent{\bf Proof: }}

\newcommand{\qed}{\hfill{\bf QED}\vspace{5mm}}

\begin{document}
\title{Simple heteroclinic cycles in $\R^4$}

\author{Olga Podvigina\\
Institute of Earthquake Prediction Theory\\
and Mathematical Geophysics\\
84/32 Profsoyuznaya St, 117997 Moscow, Russian Federation; \\
UNS, CNRS, Lab. Lagrange, OCA,\\
BP~4229, 06304 Nice Cedex 4, France;\\
and \\
Pascal Chossat \\
Laboratoire J-A Dieudonn\'e, CNRS - UNS \\
Parc Valrose \\
06108 Nice Cedex 2, France
}

\maketitle

\begin{abstract}
In generic dynamical systems heteroclinic cycles are invariant sets of codimension at
least one, but they can be structurally stable in systems which are equivariant under the action of a symmetry group,
due to the existence of flow-invariant subspaces. For dynamical systems in $\R^n$ the minimal dimension for which such robust
heteroclinic cycles can exist is $n=3$. In this case the list of admissible
symmetry groups is short and well-known. The situation is
different and more interesting when $n=4$. In this paper we list all finite
groups $\Gamma$ such that an open set of smooth $\Gamma$-equivariant
dynamical systems in $\R^4$ possess a simple
heteroclinic cycle (a structurally stable heteroclinic cycle satisfying certain additional constraints).
This work extends the results which were obtained by Sottocornola in the case when all equilibria
in the heteroclinic cycle belong to the same $\Gamma$-orbit (in this case one speaks of homoclinic cycles).
\end{abstract}

\section{Introduction}\label{sec_intro}

Heteroclinic cycles are flow-invariant sets produced by dynamical systems, which have the property to carry recurrent dynamics with
intermittent, cycling switching between equilibria (or more complicated bounded invariant sets but we shall restrict here to steady
states). These objects are known to exist and in addition to be structurally stable within certain classes of $\Gamma$-equivariant
systems, where $\Gamma$ is a finite or compact Lie group. Here we consider continuous dynamical systems
\begin{equation}\label{eq_ode}
\dot{\bf x}=f({\bf x}),\quad f:\R^n\to\R^n
\end{equation}
with the equivariance condition
\begin{equation}\label{sym_ode}
f(\gamma{\bf x})=\gamma f({\bf x})\quad\mbox{for all }
\gamma\in\Gamma\subset{\rm O}(n),\quad\Gamma\mbox{ finite}.
\end{equation}
Let $\xi_1,\ldots,\xi_m$, be a collection of (hyperbolic) saddle equilibria of
the above system and set  $\xi_{m+1}=\xi_1$. Let $W^u(\xi_j)$, resp.
$W^s(\xi_j)$, be the unstable, resp. stable manifold of $\xi_j$. Suppose that
for each $j=1,\dots,m$, $W^u(\xi_j)$ intersects $W^s(\xi_{j+1})$, then the
equilibria and their heteroclinic orbits form a {\em heteroclinic cycle}.
Heteroclinic orbits between saddles are generically destroyed by small
perturbations, hence such objects are unlikely to exist in generic systems.
They can however be structurally stable, or {\em robust}, in a restricted class of
equations, under the equivariance condition (\ref{sym_ode}) for some group $\Gamma$. Indeed this symmetry condition forces the
existence of flow-invariant subspaces, which are formed by the points in $\R^n$
fixed by isotropy subgroups of $\Gamma$. We write $\text{Fix}(\Sigma)$ for the
set of points which are fixed by $\Sigma$. This is a linear subspace of $\R^n$,
and moreover it is invariant by the flow of equation (\ref{eq_ode}). Suppose now
that there exists a collection of isotropy subgroups $\Sigma_j$ such that
$\xi_j$ is a saddle and $\xi_{j+1}$ is a sink in $\text{Fix}(\Sigma_j)$, with
the convention that $\xi_{m+1}=\xi_1$. Suppose in addition that a saddle-sink
connection exists from $\xi_j$ to $\xi_{j+1}$ in $\text{Fix}(\Sigma_j)$, then
this connection is robust against (smooth) perturbations in the class of
$\Gamma$-equivariant systems.

Many examples of robust heteroclinic cycles have been discovered and studied, especially in the context of hydrodynamical flows,
see \cite{cl2000,Kru97} for an overview.

The question which we address in this paper is the following: for which groups
$\Gamma$ do there exist dynamical systems as
above, which possess a structurally stable heteroclinic cycle? The answer to
this question depends on $n$ and we have to be more specific on this issue.

The case $n=3$ is the simplest one in which robust heteroclinic cycles can
occur and it can easily be handled. However when $n=4$ the situation is considerably more involved.
Examples of 4-dimensional heteroclinic cycles have been known and studied
because they provide "non-trivial" stability and bifurcation properties
\cite{Field}. A classification of genuinely 4-dimensional robust
{\em homoclinic} cycles was achieved by Sottocornola in \cite{sot03, sot05}.
A homoclinic cycle is a heteroclinic cycle in which all equilibria belong to
the same $\Gamma$-orbit. Sottocornola listed all finite subgroups of O(4)
for which robust homoclinic cycles exist. An outcome of his work is that one can find in $\R^4$ robust homoclinic cycles which
connect $2k$ equilibria with $k>2$ arbitrary large.

Our aim is at extending these results to robust heteroclinic cycles in $\R^4$.
A first classification of heteroclinic cycles was proposed in \cite{km95a}.
Assuming that all $P_j$'s are planes, the authors introduced the concept of
"simple" heteroclinic cycles, which were further divided into the classes
A, B and C. Although the finite groups admitting cycles of types B and C can be easily
found, the list of groups admitting type A was unknown. It is the aim of
this paper to fill the gap.
It was implicitely assumed in \cite{km95a,km04}
that simple heteroclinic cycles are such that each equilibrium in the cycle has generically only simple eigenvalues. We shall see in the next section that this is not always the case and we complete the definition of simple heteroclinic cycles accordingly.

Like in \cite{op13,sot03} our
analysis exploits the quaternionic presentation of finite subgroups of SO(4). It does
however not rely on Galois theory as in \cite{sot03} and it provides elementary proofs.

The paper is organized as follows: in Section \ref{sec2} we introduce basic notions about robust heteroclinic cycles and about the
presentation of SO(4) and O(4) with quaternions. These are the basic material which will be used in the rest of the paper. In
section \ref{sec3} the main theorems are stated and their proof is given through
a series of lemmas. The case $\Gamma\subset$\,SO(4) is considered first, then
$\Gamma\subset\,$O(4).
In theorem \ref{th1} the proofs that a subgroup $\Gamma$ admits, or
does not admit, simple heteroclinic cycles are presented only for selected
$\Gamma\subset$\,SO(4). For other subgroups of SO(4) the proofs are similar,
and therefore are omitted.
Annexes \ref{planereflections}, \ref{conjugacyclasses} and \ref{Sigmaj-Deltaj}
contain relevant informations about the geometry of finite
subgroups of SO(4).

In Section \ref{sec4} we show several examples of heteroclinic cycles in $\R^4$
and in Section \ref{sec5} we discuss the results together with some open questions.

Simple heteroclinic cycles, which are discussed in this paper, suppose the
existence of one dimensional fixed-point subspaces for the action of the group
in $\R^4$.
In annex \ref{subgroupsO4} we list finite subgroups of O(4), which act
irreducibly but do not possess such a subspace. This provides an alternative
and simple approach to a problem which was addressed by Lauterbach and
Matthews in \cite{laumat}.

\section{Background and notations}\label{sec2}

\subsection{Simple heteroclinic cycles in $\R^4$}
In this section we make precise the framework in which we look for robust heteroclinic cycles. Our notations will follow those of \cite{km95a}. \\
Let $\xi_1,\ldots,\xi_M$ be hyperbolic equilibria of the $\Gamma$-equivariant system
(\ref{eq_ode})--(\ref{sym_ode}) with stable and unstable manifolds $W^s(\xi_j)$
and $W^u(\xi_j)$, respectively. Assuming $\xi_{M+1}=\xi_1$, we denote by
$\kappa_j$, $j=1,\ldots,M$, the set of trajectories from $\xi_j$ to $\xi_{j+1}$:
$\kappa_j=W^u(\xi_j)\cap W^s(\xi_{j+1})\ne\emptyset$.

\begin{definition}\label{def4}
(i) The union of equilibria $\{\xi_1,\ldots,\xi_M\}$ and their connecting orbits $\{\kappa_1,\ldots,\kappa_M\}$, is called a \underline{heteroclinic cycle}. \\
(ii) a \underline{homoclinic cycle} is a heteroclinic cycle in which the $\xi_j$ belong to the {\em same} group orbit.
\end{definition}
We recall that the {\it isotropy group} of a point $x\in\R^n$ is the subgroup of $\Gamma$
satisfying
$$
\Sigma_x=\{\gamma\in\Gamma\ ~:~\ \gamma x=x\}.
$$
The {\it fixed-point subspace} of a subgroup $\Sigma\subset\Gamma$ is the subspace
$$
{\rm Fix}(\Sigma)=\{{\bf x}\in\R^n\ ~: \ \sigma{\bf x}={\bf x}\mbox{ for all }
\sigma\in\Sigma\}.
$$
When $\dim{\rm Fix}(\Sigma)=1$ (resp. 2) the subspace is sometimes called an {\em axis of symmetry}
(resp. a {\em plane of symmetry}). We shall use either denominations.
If a point $x$ has isotropy $\Sigma$, then the point $\gamma x$ has isotropy $\gamma\Sigma\gamma^{-1}$. There is a bijection between the $\Gamma$-orbit of a point and the conjugacy class of its isotropy subgroup in $\Gamma$. Another useful property is that the largest subgroup of $\Gamma$ which leave the subspace ${\rm Fix}(\Sigma)$ invariant is the normalizer $N(\Sigma)$ of $\Sigma$.

The following definition gives sufficient conditions for a heteroclinic cycle to persist under small enough $\Gamma$-equivariant
perturbations.
\begin{definition}\label{def5} {\bf \cite{km95a}}
The heteroclinic cycle is \underline{structurally stable (or robust)} if for any $j$, $1\le j\le M$, there exist
$\Sigma_j\subset\Gamma$ and $P_j={\rm Fix}(\Sigma_j)$ such that
\begin{itemize}
\item[(i)] $\xi_{j}$ is a sink in $P_j$;
\item[(ii)] $\xi_{j-1}$, $\xi_j$ and $\kappa_j$ belong to $P_j$.
\end{itemize}
\end{definition}
In case of a homoclinic cycle, it is enough to assume the existence of a
transformation $\gamma\in\Gamma$ such that a saddle-sink connection exists from
$\xi_1$ to $\xi_2=\gamma\xi_1$ in a fixed-point subspace $P$. Homoclinic cycles
in $\R^4$ have been classified by Sottocornola \cite{sot03}.

In all the following we use the notations $L_j=P_{j-1}\cap P_j=\rm{Fix}(\Delta_j)$.

\begin{figure}[h]
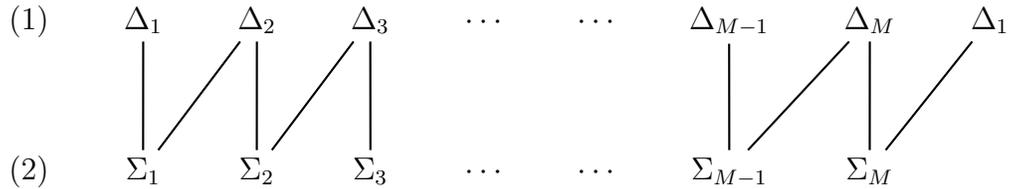

\begin{center}
$
\begin{psmatrix}[colsep=1cm]
(1) & \Delta_1 & \Delta_2 & \Delta_3  & \cdots & \cdots & \Delta_{M-1} &  \Delta_M & \Delta_1 \\ (2)
 & \Sigma_1 & \Sigma_2 & \Sigma_3 & \cdots & \cdots & \Sigma_{M-1} & \Sigma_M &
\end{psmatrix}
$
\end{center}
\ncline[nodesep=3pt]{1,2}{2,2}
\ncline[nodesep=3pt]{1,3}{2,2}
\ncline[nodesep=3pt]{1,3}{2,3}
\ncline[nodesep=3pt]{1,4}{2,3}
\ncline[nodesep=3pt]{1,4}{2,4}
\ncline[nodesep=3pt]{1,7}{2,7}
\ncline[nodesep=3pt]{1,8}{2,7}
\ncline[nodesep=3pt]{1,8}{2,8}
\ncline[nodesep=3pt]{1,9}{2,8}
\caption{\footnotesize The graph structure of the isotropy types for a simple heteroclinic cycle.
In parentheses: dimensions of the fixed-point subspaces. \label{fig:isotropylattice}}
\end{figure}

In \cite{km04} it was assumed that for all $j$, $\dim(P_j)=2$ and the heteroclinic cycle intersects each connected component of
$L_j\setminus\{0\}$ in at most one point. They called {\em simple} any robust heteroclinic cycle with these properties. Figure \ref{fig:isotropylattice} sketches the sequence of inclusions between {\em isotropy types}\footnote{An isotropy type is the conjugacy class of an isotropy subgroup.  Isotropy types are partially ordered by group inclusion, see \cite{GSS} for the introduction of this concept in bifurcation theory.} corresponding to the groups $\Sigma_i$ and $\Delta_j$ when the heteroclinic cycle is simple.

This assumption imposes constraints on the eigenvalues and eigenvectors of the Jacobian matrix $J_j=df(\xi_j)$. Because $P_j$ are flow-invariant
planes, $J_j$ has three eigenvectors which belong to respectively $L_j$, $P_{j-1}\ominus L_j$ and $P_j\ominus L_j$ where $X
\ominus Y$ denotes a complementary subspace of $Y$ in $X$.  We call {\em radial} the eigenvalue $r_j$ along the axis $L_j$, {\it
contracting} the eigenvalue $-c_j$ with eigenspace $V_j=P_{j-1}\ominus L_j$ (with $c_j>0$), {\it expanding} the eigenvalue $e_j$
with eigenspace $W_j=P_j\ominus L_j$ ($e_j>0$), {\em transverse} the remaining eigenvalue and $T_j$ the corresponding
eigenspace. Note that by construction, all eigenvalues of $J_j$ must be real.\\
We recall that the isotypic decomposition of a representation $T$ of a (finite)
group $G$ in a vector space $V$ is the decomposition
$V=V^{(1)}\oplus\cdots\oplus V^{(r)}$ where $r$ is the number of equivalence
classes of irreducible representations of $G$ in $V$ and each
$V^{(j)}=T_{|V_j}$ is the sum of the equivalent irreducible representations in
the $j$-th class. This decomposition is unique. The subspaces $V^{(j)}$ are
mutually orthogonal (if $G$ acts orthogonally).
\begin{lemma} \label{prop:isotypic decomp}
 Let a robust heteroclinic cycle in $\R^4$ be such that for all $j$:
(i) $\dim P_j=2$, (ii) each connected component of $L_j\setminus\{0\}$ is
intersected at most at one point by the heteroclinic cycle.
Then the isotypic decomposition of the
representation of $\Delta_j$ in $\R^4$ is of one of the following types:
\begin{enumerate}
\item $L_j\oplus^\perp V_j\oplus^\perp W_j\oplus^\perp T_j$ (the symbol $\oplus^\perp$ indicates the orthogonal direct sum).
\item $L_j\oplus^\perp V_j\oplus^\perp \widetilde{W_j}$ where $\widetilde{W_j}=W_j\oplus T_j$ has dimension 2.
\item $L_j\oplus^\perp \widetilde{V_j}\oplus^\perp W_j$ where $\widetilde{V_j}=V_j\oplus T_j$ has dimension 2.
\end{enumerate}
In cases 2 and 3, $\Delta_j$ acts in $\widetilde{W_j}$ (resp. $\widetilde{V_j}$) as a dihedral group $\D_m$ for some $m\geq 3$. It
follows that in case 2, $e_j$ is double (and $e_j=t_j$) while in case 3, $-c_j$ is double (and $-c_j=t_j$).
\end{lemma}
\proof $L_j$ is the axis on which $\Delta_j$ acts trivially, so it is a component of the isotypic decomposition. There can't be a 3-
dimensional component because from the existence of a heteroclinic cycle the eigenvalues of $J_j$ along $V_j$ and $W_j$ must be
of opposite signs. Therefore the remaining possibilities are that there are, in addition to $L_j$, three 1-dimensional components or
one 1-dimensional and one 2-dimensional components.
The action of $\Delta_j$ on a one dimensional component different from $L_j$ is isomorphic to $\Z_2$ (taking any non-zero vector
to its opposite). The action on a two dimensional component allows a priori more possibilities: it can be isomorphic to the $k$-fold
rotation group $C_k$ with $k\geq 3$, or to the dihedral group $\D_k$. The former case is excluded because this 2-dimensional
space must contain at least one invariant axis (and therefore at least 3 of them by the-fold rotations). Another way to prove this is
that if the action were isomorphic ot $C_m$ only, then in general the eigenvalues of $J_j$ along these components would be
complex. Hence there is a double eigenvalue, which can be either $-c_j=t_j$ or $e_j=t_j$, the corresponding isotypic component
being either $V_j$ or $W_j$.
\qed

\noindent Cases 2 and 3 of the above lemma were not accounted for in \cite{km04}.
For the sake of clarity we therefore introduce the following definition.
\begin{definition}\label{def:verysimple}
Let a robust heteroclinic cycle in $\R^4$ satisfy the conditions (i) and (ii)
of lemma \ref{prop:isotypic decomp}.
The cycle is called {\em simple} if case 1 holds true for all $j$, and
{\em pseudo-simple} otherwise.
\end{definition}

\begin{remark}
It can be easily shown that in $\R^4$ the notions of simple heteroclinic cycle
in \cite{km04} and in the above definition do coincide in the following cases:
a) the heteroclinic cycle is homoclinic;
b) the heteroclinic cycle is asymptotically stable (hence the stability analysis
for simple heteroclinic cycles in \cite{km04} is correct). \\
Also note that for simple heteroclinic cycles, $N(\Sigma_j)/\Sigma_j\cong\D_{k_j}$ where $\D_1\cong\Z_2$.
\end{remark}

In this paper we consider simple heteroclinic cycles. Pseudo-simple heteroclinic cycles will be considered in a forthcoming work.  We give in Section \ref{subsec:simple} an example of a pseudo-simple heteroclinic cycle.

The property of being simple imposes strong geometrical constraints on the symmetries allowing for a robust heteroclinic cycle. Our aim in the next sections will be to exploit these constraints in order to determine all these symmetries. For this we still need some definitions and preliminary important properties.

\begin{lemma} (see proof in \cite{km04}) \label{cor:1}
Consider a simple heteroclinic cycle in $\R^4$. For all $j$,
either $\Delta_j\cong\Z_2^2$ and $\Sigma_j\cong\Z_2$, or $\Delta_j\cong\Z_2^3$ and
$\Sigma_j\cong\Z_2^2$. Moreover the planes $P_j=\text{Fix}(\Sigma_j)$ and $P_{j+1}$ intersect orthogonally.
\end{lemma}
\begin{remark} \label{rem:planereflection}
An order two element $\sigma$ in SO(4) whose fixed point subspace is a plane
$P$ must act as $-Id$ in the plane $P^\perp$ fully perpendicular to $P$.
Nevertheless, to distinguish it from other rotations fixing the points on $P$,
we call $\sigma$ a \rm{plane reflection}.
\end{remark}

In the case $\Sigma_j\cong\Z_2$ for all $j$, the heteroclinic cycle does not
intersect with any hyperplane of symmetry (a hyperplane which is the fixed-point
subspace of some subgroup of $\Gamma$), while in the second case at least one
such hyperplane exists. Indeed if $\Sigma_j\cong\Z_2$ then
$P_j=\text{Fix}(\Sigma_j)$ can't be included in a lower isotropy proper fixed-point subspace of $\R^4$.
Based on this property, Krupa and Melbourne \cite{km04} separated heteroclinic cycles in $\R^4$ into three types.

\begin{definition}\label{def:typeABC}
A simple robust heteroclinic cycle is of \underline{type A} if $\Sigma_j\cong\Z_2$ for all $j$. It is of \underline{type B} if the heteroclinic cycle lies entirely in a fixed-point hyperplane. Otherwise it is of \underline{type C}.
\end{definition}
Krupa and Melbourne have determined in \cite{km04} the simple heteroclinic cycles of types B and C. We give this list in the following theorem, using their notations: $B^\pm_m$ indicates a heteroclinic cycle of type B with $m$ different types of equilibria (two equilibria have the same type if their isotropy groups are conjugate) and either $-I\in\Gamma$ (sign $-$) or not (sign $+$). Same notations for heteroclinic cycles of type C.
The coordinates $(x_1,x_2,x_3,x_4)$ are chosen to correspond to the isotypic decomposition of $\Delta_1$ with the trivial
component along the first coordinate. We only indicate the main features of the heteroclinic cycles, since the geometry is simple but cumbersome to describe.
\begin{theorem}[see \cite{km04}] \label{theorem:typesBC}
There are 4 different types of simple heteroclinic cycles of type B and 3 types of simple heteroclinic cycles of type C.
\begin{enumerate}
\item $B_2^+$ with $\Gamma=\Z_2^3$ consisting of the reflections
$(x_1,\pm x_2,\pm x_3, \pm x_4)$. There are three different hyperplanes and in
each of them, a heteroclinic cycle with two equilibria, one on each connected
component of $L_1\setminus\{0\}$.
\item $B_1^+$ with $\Gamma=\Z_2\ltimes\Z_2^3$ where $\Z_2^3$ acts as above and $\Z_2$ is generated by $(-
x_1,x_3,x_2,x_4)$. The structure of the heteroclinic cycle is the same as above but $\xi_1$ and $\xi_2$ are interchanged by $
\Z_2$, hence the cycle is homoclinic.
\item $B_3^-$ with $\Gamma=\Z_2^4$ generated by reflections through the four hyperplanes of coordinates. Similar heteroclinic
cycles exist in each hyperplane. For example in the hyperplane $(x_1,x_2,x_3,0)$ heteroclinic cycles connect equilibria lying on any
three axes $x_1$, $x_2$, $x_3$ and the heteroclinic connections lie in the corresponding planes of coordinates.
\item $B_1^-$ with $\Gamma=\Z_3\ltimes\Z_2^4$ where $\Z_3$ is generated by the circular permutation of $x_1,x_2,x_3$. Same as above but with all three equilibria in the same $\Z_3$-orbit, hence the cycle is homoclinic.
\item $C_4^-$ with $\Gamma=\Z_2^4$ acting as in 3. These cycles connect equilibria lying on the four coordinate axes.
\item $C_1^-$ with $\Gamma=\Z_4\ltimes\Z_2^4$ with $\Z_4$ acting by circular permutation of the coordinates. Same as above but
all equilibria in the same group orbit, hence the cycle is homoclinic.
\item $C_2^-$ with $\Gamma=\Z_2\ltimes\Z_2^4$ and $\Z_2$ generated by the permutation $(x_1,x_2)\mapsto (x_3,x_4)$. Same
as above but the 4 equilibria are pairwise of same type.
\end{enumerate}
\end{theorem}

\subsection{Quaternionic presentation of the group SO(4)}
\label{sec:quaternions}

In this section we recall some useful properties of quaternions \cite{conw,pdv}.
A real quaternion is a set of four real numbers, ${\bf q}=(q_1,q_2,q_3,q_4)$.
Introducing the elements
$i=(0,1,0,0)$, $j=(0,0,1,0)$ and $k=(0,0,0,1)$, any quaternion has the form
$q_1+q_2i+q_3j+q_4k$, where the first component is called the {\em real part} of
the quaternion. Multiplication is defined by the rules $i^2=j^2=k^2=-1$,
$ij=-ji=k$, $jk=-kj=i$, $ki=-ik=j$, which implies
\begin{equation}\label{mqua}
\begin{array}{ccc}
{\bf q}{\bf w}&=&(q_1w_1-q_2w_2-q_3w_3-q_4w_4,q_1w_2+q_2w_1+q_3w_4-q_4w_3,\\
&&q_1w_3-q_2w_4+q_3w_1+q_4w_2,q_1w_4+q_2w_3-q_3w_2+q_4w_1).
\end{array}\end{equation}
The conjugate of $\bf q$ is $\tilde{\bf q}=q_1-q_2i-q_3j-q_4k$ and
${\bf q}\tilde{\bf q}=q_1^2+q_2^2+q_3^2+q_4^2=|{\bf q}|^2$ is the square
of the norm of $\bf q$. Hence $\tilde{\bf q}$ is also the inverse ${\bf q}^{-1}$ of a unit
quaternion $\bf q$. We denote by $\mH$ the multiplicative group of unit
quaternions; obviously, its identity element is $(1,0,0,0)$.

A unit quaternion can be represented as $\bq=(\cos\theta,{\bf u}\sin\theta)$,
where ${\bf u}=(q_2,q_3,q_4)\in\R^3$ is a unit vector. The three-dimensional
subspace $w=0$ in the four-dimensional vector space of all quaternions
${\bf v}=(w,x,y,z)$ can be identified with $\R^3$. The transformation
${\bf v}\mapsto \bq{\bf v}\bq^{-1}$ is the rotation of angle $2\theta$ around
${\bf u}$ in $\R^3=\{(0,x,y,z)\}$, it is an element of SO(3). The respective homomorphism
of $\mH$ on SO(3) is 2-to-1 and its kernel is comprised of $(\pm1,0,0,0)$.

Therefore any finite subgroup of $\mH$ falls into one of the following cases,
which are pre-images of the respective subgroups of SO(3):
\begin{equation}\label{finsg}
\renewcommand{\arraystretch}{1.5}
\begin{array}{ccl}
\Z_n&=&\displaystyle{\oplus_{r=0}^{n-1}}(\cos2r\pi/n,0,0,\sin2r\pi/n)\\
\D_n&=&\Z_{2n}\oplus\displaystyle{\oplus_{r=0}^{2n-1}}(0,\cos r\pi/n,\sin r\pi/n,0)\\
\V&=&((\pm1,0,0,0))\\
\T&=&\V\oplus(\pm{1\over2},\pm{1\over2},\pm{1\over2},\pm{1\over2})\\
\mO&=&\T\oplus\sqrt{1\over2}((\pm1,\pm1,0,0))\\
\I&=&\T\oplus{1\over2}((\pm\tau,\pm1,\pm\tau^{-1},0)),
\end{array}\end{equation}
where $\tau=(\sqrt{5}+1)/2$. Double parenthesis denote all possible permutations
of quantities within the parenthesis and for $\I$ only even permutations of
$(\pm\tau,\pm1,\pm\tau^{-1},0)$ are elements of the group.
Any other finite subgroup of $\mH$ is conjugate to one of these under an inner
automorphism of $\mH$.

The 8 elements group $\V=\{(\pm1,0,0,0),\ (0,\pm1,0,0),\ (0,0,\pm1,0),\ (0,0,0,\pm1)\}$ is classically known as the quaternion group.

\medskip
The four numbers $(q_1,q_2,q_3,q_4)$ can be regarded as Euclidean coordinates
of a point in $\R^4$. For any pair of unit quaternions $({\bf l};{\bf r})$,
the transformation ${\bf q}\to{\bf lqr}^{-1}$ is a rotation in $\R^4$, i.e.
an element of the group SO(4). The mapping
$\Phi:\mH\times\mH\to$\,SO(4) that relates the pair $({\bf l};{\bf r})$
with the rotation ${\bf q}\to{\bf lqr}^{-1}$ is a homomorphism onto,
whose kernel consists of two elements, $(1;1)$ and $(-1;-1)$; thus
the homomorphism is two to one.

Therefore, a finite subgroup of SO(4) is a subgroup of a product of two
finite subgroups of $\mH$. There is however an additional subtlety.
Let $\Gamma$ be a finite subgroup of SO(4), ${\cal G}=\Phi^{-1}(\Gamma)$ and $({\bf l}_j;{\bf r}_j)$,
$1\le j\le J$, its elements. Denote by $\bf L$ and $\bf R$ the finite subgroups of $\mH$
generated by ${\bf l}_j$ and ${\bf r}_j$, $1\le j\le J$, respectively.
To any element ${\bf l}\in \bf L$ there are several corresponding elements
${\bf r}_i$, such that $({\bf l};{\bf r}_i)\in\mH$, and similarly for any
${\bf r}\in{\bf R}$. This establishes a correspondence between $\bf L$ and
$\bf R$. Denote by ${\bf L}_K$ and ${\bf R}_K$ the subgroups of $\bf L$ and
$\bf R$ corresponding to the unit elements in $\bf R$ and $\bf L$, respectively.
The groups ${\bf L}/{\bf L}_K$ and ${\bf R}/{\bf R}_K$ are isomorphic
\cite{pdv} and characterize the group $\cal G$. Moreover,
${\cal G}_k={\bf L}_K\times {\bf R}_K$ is normal in $\cal G$ and ${\bf L}/{\bf L}_K\cong {\cal G}/{\cal G}_K$. This relation allows to compute the order of $\cal G$ and $\Gamma$ from the knowledge of ${\bf L},{\bf L}_K$ and ${\bf R}_K$.

\medskip
\noindent {\bf Notation.} Following \cite{pdv} we write
$({\bf L}\rl{\bf L}_K;{\bf R}\rl{\bf R}_K)$ for the group $\Gamma$.

\medskip
The isomorphism between ${\bf L}/{\bf L}_K$ and ${\bf R}/{\bf R}_K$ may not be
unique and different isomorphisms give rise to different subgroups of SO(4). For
instance, the correspondence
$${\bf p}^j{\bf r}\leftrightarrow{\bf p}^{'sj}{\bf r}',$$
where ${\bf r}\in\Z_{mr}/\Z_m$, ${\bf r}'\in\Z_{nr}/\Z_n$,
${\bf p}=(\cos2\pi/mr,0,0,\sin2\pi/mr)$ and\\
${\bf p}'=(\cos2\pi/nr,0,0,\sin2\pi/nr)$, for different $s<r/2$ prime to $r$,
gives geometrically distinct
subgroups of SO(4), which are denoted by $(\Z_{mr}\rl\Z_m;\Z_{nr}\rl\Z_n)_s$.
The isomorphism extended to the one between $\D_{mr}/\Z_m$ and $\D_{nr}/\Z_n$
defines a group $(\D_{mr}\rl\Z_m;\D_{nr}\rl\Z_n)_s$.
The isomorphism between $\mO/\Z_1$ and $\mO/\Z_1$ can be the identity, or
it can be ${\bf r}={\bf l}$ for ${\bf r}\in\T$ and ${\bf r}=-{\bf l}$  for
${\bf r}\in\T_1$, where $\T_1$ is the coset of $\T$ in $\mO$. The latter
subgroup is denoted $(\mO\rl\Z_1;\mO\rl\Z_1)^{\dagger}$.
The complete list of finite subgroups of SO(4) is given in table \ref{listSO4}.

Here we are interested in subgroups $\Gamma$ of SO(4) such that
a $\Gamma$-equivariant system can possess a heteroclinic cycle. As it will be
shown in lemma \ref{lem5}, a preimage
$\Phi^{-1}\Gamma=({\bf L}\rl{\bf L}_K;{\bf R}\rl{\bf R}_K)$ must satisfies
$\D_2\subset{\bf L}$ and $\D_2\subset{\bf R}$.
The subgroups of SO(4) where both $\bf L$ and $\bf R$ contain $\D_n$ ($n>1$) are
the groups 10-32 and 34-39 in the table.

\begin{table}[htdp]
\hskip -1cm\begin{tabular}{|c|c|c|c|c|c|c|c|c|c|c|}
\hline \# & group & order && \# & group & order && \# & group & order \\ \hline
1 & $(\Z_{2nr}\rl\Z_{2n};\Z_{2kr}\rl\Z_{2k})_s$ & $2nkr$ &&
15 & $(\D_n\rl\D_n;\mO\rl\mO)$ & $96n$ &&
29 & $(\mO\rl\mO;\I\rl\I)$ & 2880  \\ \hline

2 & $(\Z_{2n}\rl\Z_{2n};\D_{k}\rl\D_{k})_s$ & $4nk$ &&
16 & $(\D_n\rl\Z_{2n};\mO\rl\T)$ & $48n$ &&
30 & $(\I\rl\I;\I\rl\I)$ & 7200 \\ \hline

3 & $(\Z_{4n}\rl\Z_{2n};\D_{k}\rl\Z_{2k})$ & $4nk$ &&
17 & $(\D_{2n}\rl\D_n;\mO\rl\T)$ & $96n$ &&
31 & $(\I\rl\Z_2;\I\rl\Z_2)$ & 120 \\ \hline

4 & $(\Z_{4n}\rl\Z_{2n};\D_{2k}\rl\D_{k})$ & $8nk$ &&
18 & $(\D_{3n}\rl\Z_{2n};\mO\rl\V)$ & $48n$ &&
32 & $(\I^{\dagger}\rl\Z_2;\I\rl\Z_2)$ & 120 \\ \hline

5 & $(\Z_{2n}\rl\Z_{2n};\T\rl\T)$ & $24n$ &&
19 & $(\D_n\rl\D_n;\I\rl\I)$ & $240n$ &&
33 & $(\Z_{2nr}\rl\Z_{n};\Z_{2kr}\rl\Z_{k})_s$ & nkr \\ \cline{1-7}

6 & $(\Z_{6n}\rl\Z_{2n};\T\rl\V)$ & $24n$ &&
20 & $(\T\rl\T;\T\rl\T)$ & 288 &&
   & $n\equiv k\equiv 1$ (mod 2) & \\ \hline

7 & $(\Z_{2n}\rl\Z_{2n};\mO\rl\mO)$ & $48n$ &&
21 & $(\T\rl\Z_2;\T\rl\Z_2)$ & 24 &&
34 & $(\D_{nr}\rl\Z_{n};\D_{kr}\rl\Z_{k})_s$ & 2nkr \\ \cline{1-7}

8 & $(\Z_{2n}\rl\Z_{2n};\mO\rl\T)$ & $48n$ &&
22 & $(\T\rl\V;\T\rl\V)$ & 96 &&
   & $n\equiv k\equiv 1$ (mod 2) & \\ \hline

9 & $(\Z_{2n}\rl\Z_{2n};\I\rl\I)$ & $120n$ &&
23 & $(\T\rl\T;\mO\rl\mO)$ & 576 &&
35 & $(\T\rl\Z_1;\T\rl\Z_1)$ & 12 \\ \hline

10 & $(\D_n\rl\D_n;\D_k\rl\D_k)$ & $8nk$ &&
24 & $(\T\rl\T;\I\rl\I)$ & 1440 &&
36 & $(\mO\rl\Z_1;\mO\rl\Z_1)$ & 24 \\ \hline

11 & $(\D_{nr}\rl\Z_{2n};\D_{kr}\rl\Z_{2k})_s$ & $4nkr$ &&
25 & $(\mO\rl\mO;\mO\rl\mO)$ & 1152 &&
37 & $(\mO\rl\Z_1;\mO\rl\Z_1)^{\dagger}$ & 24 \\ \hline

12 & $(\D_{2n}\rl\D_n;\D_{2k}\rl\D_k)$ & $16nk$ &&
26 & $(\mO\rl\Z_2;\mO\rl\Z_2)$ & 48 &&
38 & $(\I\rl\Z_1;\I\rl\Z_1)$ & 60 \\ \hline

13 & $(\D_{2n}\rl\D_n;\D_k\rl\Z_{2k})$ & $8nk$ &&
27 & $(\mO\rl\V;\mO\rl\V)$ & 192 &&
39 & $(\I^{\dagger}\rl\Z_1;\I\rl\Z_1)$ & 60 \\ \hline

14 & $(\D_n\rl\D_n;\T\rl\T)$ & $48n$ &&
28 & $(\mO\rl\T;\mO\rl\T)$ & 576 && &&\\ \hline

\end{tabular}
\caption{Finite subgroups of SO(4)}\label{listSO4}
\end{table}
The superscript $\dagger$ is employed to denote subgroups of SO(4) where the
isomorphism between the quotient groups ${\bf L}/{\bf L}_K$ and
${\bf R}/{\bf R}_K\cong{\bf L}/{\bf L}_K$ is not the identity. The group
$\I^{\dagger}$, isomorphic to $\I$, involves the elements
$((\pm\tau^*,\pm1,\pm(\tau^*)^{-1},0))$, where $\tau^*=(-\sqrt{5}+1)/2$.
The groups 1-32 contain the central rotation $-I$, and the groups 33-39 do not.

\medskip
A reflection in $\R^4$ can be expressed in the quaternionic presentation as
${\bf q}\to{\bf a\tilde qb}$, where ${\bf a}$ and ${\bf b}$ is a pair of
unit quaternions. We write this reflection as $({\bf a};{\bf b})^*$.
The transformations ${\bf q}\mapsto {\bf a\tilde qa}$ and ${\bf q}\mapsto -{\bf a\tilde qa}$ are respectively the reflections about the axis $\bf a$ and through the hyperplane orthogonal to the vector $\bf a$. Therefore if $\bf a\perp b$ are two orthogonal unit quaternions, the rotation of angle $\pi$ about the plane $<{\bf a},{\bf b}>$ is ${\bf q}\to-{\bf a\tilde bq}({\bf\tilde ba)}$. We call this transformation the {\em plane reflection} about $<{\bf a},{\bf b}>$ (see Remark \ref{rem:planereflection}).

A group $\Gamma^*\subset$\,O(4), $\Gamma^*\not\subset$\,SO(4), can be decomposed as
$$\Gamma^*=\Gamma\oplus\sigma\Gamma,\hbox{ where $\Gamma\subset$\,SO(4)
and $\sigma=({\bf a};{\bf b})^*\notin$\,SO(4)}.$$
If $\Gamma^*$ is finite, then in the quaternion form of $\Gamma$,
$\Phi^{-1}\Gamma=({\bf L}\rl{\bf L}_K;{\bf R}\rl{\bf R}_K)$, the groups ${\bf L}$
and ${\bf R}$ are isomorphic, and so are ${\bf L}_K$ and ${\bf R}_K$ \cite{pdv}.
The elements ${\bf a}$ and ${\bf b}$ belong to a subgroup $\bf H$ of $\mH$ in
which ${\bf G}={\bf L}={\bf R}$ and ${\bf G}_K={\bf L}_K={\bf R}_K$ are
invariant subgroups. Moreover, ${\bf a}$ and ${\bf b}$ are in the same coset of
${\bf G}$ in $\bf H$.
If $\phi$ denotes the isomorphism between ${\bf L'}={\bf L}/{\bf L}_K$ and
${\bf R'}={\bf R}/{\bf R}_K$, $\alpha$ and $\beta$ the isomorphisms from
${\bf R'}$ to ${\bf L'}$ defined by
$\alpha:{\bf R'}\to{\bf a}{\bf R'}{\bf a}^{-1}$ and
$\beta:{\bf R'}\to{\bf a}{\bf R'}{\bf a}^{-1}$ then $\phi\alpha\phi\beta=1$.
The list of finite subgroups of O(4), which was derived from these arguments,
can be found in \cite{pdv}.

\section{Classification of simple heteroclinic cycles in $\R^4$}
\label{sec3}

In this section we state and prove classification of simple
heteroclinic cycles in $\R^4$. More precisely, we list all finite subgroups
$\Gamma\subset$\,O(4) such that $\Gamma$-equivariant systems
exist, which possess a simple heteroclinic cycle. Note,
that the subgroups $\Gamma$ giving rise to cycles of types B and C were found
in \cite{km04} and are listed in Theorem \ref{theorem:typesBC} (see previous section). \\
The proof of our main theorems is given in Section \ref{mainproofs} and will proceed
from a series of lemmas which are stated in Section \ref{lems}.

\subsection{Statement of the main results}

We begin with a definition.
\begin{definition} \label{def:admits}
We say that a subgroup $\Gamma$ of O($n$) {\em admits} robust heteroclinic cycles if there exists an open subset of the set of smooth $\Gamma$-equivariant vector fields in $\R^n$, such that vector fields in this subset possess a (robust) heteroclinic cycle.
\end{definition}
The following theorems exhibit all finite subgroups of O(4), which admit robust simple heteroclinic cycles. In Theorem \ref{th1} we list those groups which are included in SO(4) and in Theorem \ref{th2} we list the groups which contain elements not in SO(4).
We use the notations introduced in \ref{sec:quaternions}.

\pagebreak

\begin{theorem}\label{th1}
A group $\Gamma\subset$S\,O(4) admits simple heteroclinic cycles,
if and only if it is one of the following:

\hskip -2cm\begin{table}[h]
\begin{equation}\label{listth1}
\renewcommand{\arraystretch}{1.5}
\begin{array}{|l|l|l|}
\hline
(\D_{2K_1}\rl\D_{2K_1};\D_{2K_2}\rl\D_{2K_2}) &
(\D_{2K}\rl\Z_{4K};\mO\rl\T),\ K\ne3k \\
\hline
(\D_{2K_1r}\rl\Z_{4K_1};\D_{2K_2r}\rl\Z_{4K_2})_s,\ K_1,K_2,r,s\hbox{ satisfy (\ref{condno1})}&
(\D_{2K}\rl\D_K;\mO\rl\T),\ K\ne2(2k+1)\\
\hline
(\D_{2K_1r}\rl\Z_{2K_1};\D_{2K_2r}\rl\Z_{2K_2})_s,\ K_1,K_2\hbox{ odd,}&
(\D_{2K}\rl\D_{2K};\I\rl\I),\ K\ne5k\\
\cline{2-2}
K_1,K_2,r,s\hbox{ satisfy (\ref{condno2})}&
(\T\rl\Z_2;\T\rl\Z_2)\\
\hline
(\D_{2K_1}\rl\D_{K_1};\D_{2K_2}\rl\D_{K_2})&
(\T\rl\T;\mO\rl\mO) \\
\hline
(\D_{2K_1}\rl\D_{K_1};\D_{2K_2}\rl\Z_{4K_2}),\ K_1\hbox{ even, }K_1/2,K_2\hbox{ co-prime} &
(\mO\rl\mO;\I\rl\I) \\
\hline
(\D_{2K_1}\rl\D_{K_1};\D_{2K_2}\rl\Z_{4K_2}),\ K_1\hbox{ odd}&
(\D_{2K_1r}\rl\Z_{K_1};\D_{2K_2r}\rl\Z_{K_2})_s,\
K_1,K_2\hbox{ odd,}\\
\cline{1-1}
(\D_{2K}\rl\D_{2K};\T\rl\T)&
K_1,K_2,r,s\hbox{ satisfy (\ref{condno3})}\\
\hline
(\D_{2K}\rl\D_{2K};\mO\rl\mO),\ K\hbox{ odd} &\\
\hline
\end{array}
\end{equation}
\end{table}
\end{theorem}

\begin{remark}
A subgroup $\Gamma\subset$\,O($n$) was called in \cite{sot03,sot05} a minimal
admissible group if
\begin{itemize}
\item $\Gamma$ admits simple homoclinic cycles;
\item any proper subgroup of $\Gamma$ does not admit homoclinic cycles.
\end{itemize}
Minimal admissible groups, subgroups of O(4), were found in
\cite{sot03,sot05}.
In the quaternion form subgroups of SO(4) are $(\D_4\rl\D_2;\D_4\rl\Z_8)$
(with $\alpha=\pi/2$) and $(\D_4\rl\D_4;\T\rl\T)$ (with $\alpha=\pi/4$). Any
group admitting simple homoclinic cycles (see the table in annex \ref{Sigmaj-Deltaj}),
except for $(\T\rl\Z_2;\T\rl\Z_2)$, has at least one of
these groups as a subgroup. A homoclinic cycle which can exist in a
$(\T\rl\Z_2;\T\rl\Z_2)$-equivariant system belongs to a three-dimensional
hyperplane, such cycles were not considered {\it ibid}.
\end{remark}

\pagebreak

\begin{theorem}\label{th2}
A group $\Gamma^*\subset$\,O(4),
$$\Gamma^*=\Gamma\oplus\sigma\Gamma,
\hbox{ where $\Gamma\subset$\,SO(4) and $\sigma\notin$\,SO(4)},$$
admits simple heteroclinic cycles, if and only if
${\Gamma}$ and $\sigma$ are one of the following:
\begin{table}[h]
\begin{equation}\label{listth2}
\renewcommand{\arraystretch}{1.5}
\begin{array}{|l|l|l|}
\hline
{\Gamma} & \sigma \\
\hline
(\D_4\rl\Z_2;\D_4\rl\Z_2) & ((0,1,0,1),(0,1,0,1))^*/2 \\
\hline
(\D_4\rl\Z_1;\D_4\rl\Z_1)_3 & ((0,1,0,1),(0,1,0,1))^*/2 \\
\hline
(\D_2\rl\Z_2;\D_2\rl\Z_2) & ((0,1,0,0),(0,1,0,0))^* \\
\hline
(\T\rl\Z_2;\T\rl\Z_2) & ((0,1,0,0),(0,1,0,0))^* \\
\hline
(\D_2\rl\Z_1;\D_2\rl\Z_1) & ((1,0,0,0),(1,0,0,0))^* \\
\hline
(\D_{2K}\rl\D_K;\D_{2K}\rl\D_K) & ((\cos\theta,0,0,\sin\theta),(1,0,0,0))^*,\ \theta=\pi/(2K) \\
\hline
\end{array}
\end{equation}
\end{table}
\end{theorem}

\begin{remark}
The groups listed in Theorem \ref{theorem:typesBC}, which admit heteroclinic cycles of
types B or C, are not subgroups of SO(4) and therefore decompose as
$\Gamma^*=\Gamma\oplus\sigma\Gamma$, $\Gamma\subset$\,SO(4).
In quaternion formulation the groups $\Gamma$ are the following: \\
$(\D_2\rl\Z_1;\D_2\rl\Z_1)$ (for $B_2^+$);
$(\D_4\rl\Z_1;\D_4\rl\Z_1)_3$ (for $B_1^+$);
$(\D_2\rl\Z_2;\D_2\rl\Z_2)$ (for $B_3^-$ and $C_4^-$);
$(\T\rl\Z_2;\T\rl\Z_2)$ (for $B_1^-$);
$(\D_2\rl\D_2;\D_2\rl\D_2)$ (for $C_2^-$);
$(\D_4\rl\D_2;\D_4\rl\D_2)$ (for $C_1^-$).
\end{remark}

\begin{remark}
There exists only one group $\Gamma^*\subset$\,O(4),
$\Gamma^*\not\subset$\,SO(4), admitting homoclinic cycles which are not of
type B or C \cite{sot03,sot05}. In the quaternion form its rotation subgroup
is $(\D_{2K}\rl\D_K;\D_{2K}\rl\D_K)$.
\end{remark}

These theorems are proven in Section \ref{mainproofs}, but we need first several lemmas which are provided in the next section.

\subsection{Lemmas}
\label{lems}

According to definition \ref{def:verysimple}, if $X$ is a simple heteroclinic cycle
then $\dim P_j=2$ and the plane $P_j$ intersects with $P_{j+1}$ orthogonally for
any $j$. Denote by $P_j^{\perp}$ the orthogonal complement to $P_j$ in $\R^4$.
We assume that the bases $({\bf h}_1,{\bf h}_2)$ in $P_j$ and
$({\bf h}_3,{\bf h}_4)$ in $P_j^{\perp}$ constitute a positively oriented basis
$({\bf h}_1,{\bf h}_2,{\bf h}_3,{\bf h}_4)$ in $\R^4$.
The plane $P_j^{\perp}$ intersects orthogonally with $P_{j-1}$ and $P_{j+1}$.

\begin{definition}\label{defsa}
Denote by $\alpha_j$ the oriented angle between $L_j$ and $L_{j+1}$; by $\beta_j$
the oriented angle between intersections of $P_j^{\perp}$ with $P_{j-1}$ and
$P_{j+1}$. The angles $\alpha_j$ and $\beta_j$, $1\le j\le M$, are called the
\underline{structure angles} of the heteroclinic cycle $X$.
\end{definition}

\begin{remark}
The structure angles can be alternatively defined as the angles between: (i) the expanding
eigenvector of $df(\xi_j)$ and the contracting eigenvector of $df(\xi_{j+1})$ (the angle $\alpha_j$);
(ii) the contracting eigenvector of $df(\xi_j)$ and the expanding eigenvector of
$df(\xi_{j+1})$ (the angle $\beta_j$).
\end{remark}

\begin{remark}
The definition of structure angles can be generalized to simple heteroclinic
cycles in $\R^n$ by introducing subspaces $U_j=\R^4$ such that
$P_s\subset U_j$ for $s=j-1,j$ and $j+1$.
\end{remark}

\begin{lemma} (see proof in \cite{op13})\label{lemn6}
Let $N_1$ and $N_2$ be two planes in $\R^4$ and $p_j$, $j=1,2$, be
the elements of SO(4) which act on $N_j$ as identity, and on $N_j^{\perp}$ as $-I$,
and $\Phi^{-1}p_j=({\bf l}_j;{\bf r}_j)$, where $\Phi$ is the homomorphism
defined in the previous section. Denote by $({\bf l}_1{\bf l}_2)_1$
and $({\bf r}_1{\bf r}_2)_1$ the first components of the respective quaternion
products. The planes $N_1$ and $N_2$ intersect if and only if
$({\bf l}_1{\bf l}_2)_1=({\bf r}_1{\bf r}_2)_1=\cos\alpha$ and $\alpha$ is
the angle between the planes.
\end{lemma}

In order to insure the existence of a heteroclinic cycle in terms of Definition \ref{def4}, it is enough to find $m\leq M$ and $\gamma\in\Gamma$ such that a minimal sequence of robust heteroclinic connections $\xi_1\rightarrow\cdots\rightarrow\xi_{m+1}$ exists with $\xi_{m+1}=\gamma\xi_1$ (minimal in the sense that no other equilibrium inside this sequence belongs to the $\Gamma$-orbit of $\xi_1$). It follows that $\gamma^k=1$ where $k$ is a divisor of $M$.
\begin{definition}\label{def42}
The sequence $\xi_1\rightarrow\cdots\rightarrow\xi_{m}$ and the element $\gamma$ define a {\em building block} of the heteroclinic cycle
\end{definition}

\begin{lemma}\label{lem0}
Let $\xi_1\to\ldots\to\xi_m$, $m\ge2$, be a building block of a simple
heteroclinic cycle in $\R^n$ and $\alpha_j=\pi/k_j$ be its structure angles
according to definition \ref{defsa}. Then
\begin{itemize}
\item[(a)] One of the following takes place:
\begin{itemize}
\item[(i)] all $k_j$ are even and $\Delta_i$ and $\Delta_j$ are not
conjugate for any $1\le i,j\le m$, $i\ne j$;
\item[(ii)] $m=2$, $k_1$ and $k_2$ are odd and $\Delta_1$, $\Delta_2$ are
conjugate. The case $k_j=1$ corresponds to having only one axis $L_j$ in $P_j$.
\end{itemize}
\item[(b)] The groups $\Sigma_i$ and $\Sigma_j$ are not conjugate for any
$1\le i,j\le m$, $i\ne j$.
\end{itemize}
\end{lemma}

\proof
We start from proving that either all $k_j$ are odd, or all $k_j$ are even.
Suppose that this is not true and there exists $j$ such that $k_{j-1}$ is odd
and $k_j$ is even. Denote the two connected components of $L_j\setminus\{0\}$ by
$L_j'$ and $L_j''$ and assume that $\xi_j\in L_j'$. Since $k_{j-1}$ is odd,
$\Delta_{j-1}$ and $\Delta_j$ are conjugate by some
$\kappa\in N(\Sigma_{j-1})/\Sigma_{j-1}\cong \D_{k_{j-1}}$. The symmetry
$\kappa$ satisfies $\kappa L_{j-1}=L_j$ and
$\kappa\xi_{j-1}\in L_j''$. Since $k_j$ is even, there
exists $\sigma\in N(\Sigma_j)/\Sigma_j$, such that $\sigma L_j''=L_j'$.
The image of $\xi_{j-1}$ under $\sigma\kappa$ satisfies
$\sigma\kappa\xi_{j-1}\in L_j'$, which contradicts the condition $m\ge 2$.
Hence, either all $k_j$ are even or all $k_j$ are odd.

\medskip
(a.i) Let all $k_j$ be even and assume that $\Delta_i$ and $\Delta_j$, $i\ne j$,
are conjugate by some $\sigma\in\Sigma$, which implies $\sigma L_i=L_j$.
Denote by $L_j'$ the connected component of $L_j\setminus\{0\}$ such that
$\xi_j\in L_j'$ and $L_j''=L_j\setminus\{0\}\setminus L_j'$. Since $k_j$ is
even, there exists $\kappa\in\Sigma$ such that $\kappa L_j'=L_j''$. Hence,
either $\sigma\xi_i\in L_j'$ or $\kappa\sigma\xi_i\in L_j'$. Therefore,
the assumption that $\Delta_i$ and $\Delta_j$ are conjugate contradicts
definition \ref{def42}.

\medskip
(a.ii) If all $k_j$ are odd, then there exist $\kappa_1$ and $\kappa_2$ such that
$\kappa_1L_1'=L_2''$ and $\kappa_2L_2''=L_3'$ (as above, $\xi_j\in L_j'$ for
$j=1,2,3$). Therefore, $\kappa_2\kappa_1\xi_1\in L_3'$, which implies $m\le2$.

\medskip
(b) If all $k_j$ are even, then conjugacy of $\Sigma_i$ and $\Sigma_j$ implies
that $\Delta_i$ is conjugate to $\Delta_j$ or to
$\Delta_{j+1}$, which is not possible due to (a.i).

If all $k_j$ are odd and $m=2$, then conjugacy of $\Sigma_1$ and $\Sigma_2$
and definition of the building block implies existence of $\sigma\in\Gamma$,
such that $\sigma\Sigma_1\sigma^{-1}=\Sigma_2$ and $\sigma\xi_2=\xi_2$. Hence,
the symmetry $\sigma$ maps the connection $\xi_1\to\xi_2\subset\Sigma_1$ to
the one $\xi_3\to\xi_2\subset\Sigma_2$, $\xi_3=\gamma\xi_1\gamma^{-1}$, while
the connection $\xi_2\to\xi_3$ is needed to complete the heteroclinic cycle.
\qed

In the next lemma we list the conditions for a finite subgroup of O(4) to {\em admit} (see Definition \ref{def:admits}) simple heteroclinic cycles. This lemma generalizes to heteroclinic cycles a theorem which was stated for homoclinic cycles in \cite{am02} (Theorem 4.1).

\begin{lemma}\label{lem12}
A finite subgroup $\Gamma$ of O($n$) {\em admits} simple heteroclinic cycles in $\R^n$ (see definition
\ref{def:admits}) if and only if there exist two
sequences of isotropy subgroups $\Sigma_j$, $\Delta_j$, $j=1,\dots, m$, and an
element $\gamma$ in $\Gamma$ satisfying the following conditions:
\begin{itemize}
\item[\bf{C1}.] Denote $P_j=\Fix(\Sigma_j)$ and
$L_j=\Fix(\Delta_j)$. Then $\dim P_j=2$ and $\dim L_j=1$ for all $j$.
\item[\bf{C2}.] For $i\neq j$, $\Sigma_i$ and $\Sigma_j$ are not conjugate.
\item[\bf{C3}.] For $j=2,\dots,m$, $L_j=P_{j-1}\cap P_j$, and
$L_1=\gamma^{-1}P_m\gamma\cap P_1$. We set $\Delta_{m+1}=\gamma\Delta_1\gamma^{-1}$.
\item[\bf{C4}.] $N(\Sigma_{j})/\Sigma_{j}\cong \D_{k_{j}}$, the dihedral
group of order $2k_j$. Either all $k_j$ are even and the groups $\Delta_i$,
$\Delta_j$ are not conjugate, or all $k_j$ are odd and $m\le2$.
Moreover any isotropy subgroup which contains $\Sigma_j$ is conjugate to either $\Delta_j$ or $\Delta_{j+1}$ (for any $j=1,\dots,m$).
\item[\bf{C5}.] For all $j$, the subspaces
$L_j$, $P_{j-1}\ominus L_j$ and $P_j\ominus L_j$ are one-dimensional
isotypic components in the isotypic decomposition of $\Delta_j$ in $\R^n$.
\end{itemize}
\end{lemma}
\proof
We prove sufficiency. Necessity follows from the definition of a simple heteroclinic cycle and the fact that if an invariant axis exists in $P_j$, which is not an axis of symmetry of $\D_{k_{j}}$, then its orthogonal complement in $P_j$ can't be an isotypic component for the action of $\Delta_j$ (hence a heteroclinic cycle involving a connection in $P_j$ with that axis can't be simple). \\
Hypothesis {\bf C3} results in the following property of the invariant subspaces:
for $j=2,\dots,m$, $L_{j}=P_{j-1}\cap P_{j}$, and $\gamma L_1=P_m\cap\gamma P_1$
(which also means that $L_1=\gamma^{-1}P_m\cap P_1$). Condition {\bf C4} takes
care of the case when the heteroclinic cycle connects equilibria which have the
same isotropy type. In this case the building block reduces to two equilibria. \\
Now let $X_1$ be the set of $\Gamma$-equivariant smooth vector fields in $\R^n$
which have an hyperbolic equilibrium $\xi_j$ with isotropy $\Delta_j$ for all
$j=1,\dots,m$, and such that the linearization at $\xi_j$ has a negative
eigenvalues along $L_j=\text{Fix}(\Delta_j)$, a positive eigenvalue in
$P_j=\text{Fix}(\Sigma_j)$ (in the direction orthogonal to $L_j$) and a negative
eigenvalue in $P_{j-1}=\text{Fix}(\Sigma_{j-1})$ (in the direction orthogonal to
$L_j$). Condition {\bf C3} implies that $\gamma L_1\subset P_m$ and we assume
that $\gamma\xi_1$ has a negative eigenvalue in $P_m$ in the direction
orthogonal to $\gamma L_1$. This set is non-empty and open in the space of
$\Gamma$ equivariant, smooth vector fields in $V=\R^n$.

Let $X_2$ be the set of vector fields in $X_1$ such that for all $j$, a
heteroclinic orbit connecting $\xi_j$ to $\xi_{j+1}$ exists in $P_j$ (we set
$\xi_{m+1}=\gamma\xi_1$). Since these trajectories realize saddle-sink
connections in invariant subspaces $P_j$, the set $X_2$ is open. We need to show it is not empty.  \\
Let $V=\R^n$. We need recall first some properties of the orbit space $V/\Gamma$ of a finite group action, see \cite{koenig,cl2000} for details. The orbit space can be realized as the image of the map $\Pi:V\rightarrow\R^p$, which to any point $x$ associates $(\theta_1(x),\dots,\theta_p(x))$ where $\theta_1,\dots,\theta_p$ are a (minimal) generating family of the ring of $\Gamma$-invariant polynomials in $V$. The set $\Pi(V)$ is a stratified semi-algebraic set. Each stratum is an algebraic manifold, image under $\Pi$ of the set of points in $V$ which have the same orbit type (that is, points which have conjugate isotropy subgroups). Despite the fact that $V/\Gamma$ is not a manifold we can give a meaning to a "smooth" vector field in $V/\Gamma$ by saying that it is the restriction to $\Pi(V)$ of a smooth vector field in $\R^p$, which in addition is tangent to each stratum in $\Pi(V)$. The projection of a smooth $\Gamma$-equivariant vector field in $V$ under $\Pi$ is a smooth
 vector field in $V/\Gamma$. Conversely, any smooth vector field in $V/\Gamma$ lifts to a smooth $\Gamma$-equivariant vector field in $V$ \cite{schwarz}. Another important property of the orbit space is that given a point $x\in V$ with isotropy $\Sigma$, there exists a neighborhood of $x$ in which $V/\Gamma$ is isomorphic to a neighborhood of 0 in $V/\Sigma$.

Now let $\tilde\xi_j$ be the image in $V/\Gamma$ of the equilibria $\xi_j$ for
a vector field $f$ in $X_1$ and let $\tilde f$ be the image of $f$ in $\R^p$.
We call $\tilde L_j$ the Jacobian matrix of $\tilde f$ at $\tilde\xi_j$. We also
write $S_j$ the stratum corresponding to the orbit type of the subgroup
$\Sigma_j$. Note that dim$(S_j)=2$. It follows from the properties of the orbit
space studied in \cite{koenig} that the unstable manifold of $\tilde\xi_j$
intersects $S_j$ along a one dimensional curve $w_j$ while the stable manifold
of $\tilde\xi_{j+1}$ contains a neighborhood of this point in $S_j$. Due to
second part of the condition {\bf C4} one can build a smooth path in $S_j$ which
joins $\tilde\xi_j$ and $\tilde\xi_{j+1}$ and coincides with $w_j$ in the
vicinity of $\tilde\xi_j$. The union of these paths for $j=1$ to $p$ is a closed
path $C$. Taking a tubular neighborhood of $C$ in $\R^p$ we can build a smooth
vector field $\tilde f$ which vanishes outside this neighborhood, coincides with
$\tilde L_j$ in a neighborhood of $\xi_j$, is tangent to the strata in $\Pi(V)$
and such that the unstable manifold at $\tilde\xi_j$ intersects the stable
manifold at $\tilde\xi_{j+1}$ in $S_j$ (see \cite{am02} for details). This
vector field lifts to a $\Gamma$-equivariant vector field in $V$, which belongs to $X_2$. \\
Finally the assumption {\bf C5} insures that the heteroclinic cycles are simple.
\qed

\begin{lemma}\label{lem6}
Let $P_1$ and $P_2$ be two-dimensional planes in $\R^n$,
$\dim(P_1\cap P_2)=1$, $\rho\in$\,O($n$) is a plane reflection about
$P_1$ and $\sigma\in$\,O($n$) maps $P_1$ into $P_2$. Suppose
that $\rho$ and $\sigma$ are elements of a finite subgroup
$\Delta\subset$\,O($n$). Then $\Delta\supset\D_m$, where $m\ge 3$.
\end{lemma}
\proof
Let $\bf e_1$ denote a vector in $P_1$, which is orthogonal to $P_1\cap P_2$.
According to the statement of the lemma, $\sigma^l{\bf e_1}={\bf e_1}$ for
a finite $l$. The subspace of $\R^n$ spanned by
${\bf e_1},\sigma{\bf e_1},\ldots,\sigma^{l-1}{\bf e_1}$ has at least one
$\sigma$- and $\rho$-invariant plane, which can not be decomposed as a sum
of two one-dimensional invariant subspaces. The action of group generated by
$\rho$ and $\sigma$ on this plane is isomorphic to a dihedral group $\D_k$
for a $k>2$.
\qed

\begin{lemma}\label{lem5}
Let $X$ be a simple heteroclinic cycle in a $\Gamma$-equivariant
system (\ref{eq_ode})--(\ref{sym_ode}) in $\R^4$ and $\alpha_j$ and $\beta_j$,
$j=1,\ldots,m$, be its structure angles.
Denote by $s_j$ the plane reflection through $P_j$. Then
\begin{itemize}
\item[(i)] $\alpha_j=\pi/K_j$, $K_j\in\Z$;
\item[(ii)] $\beta_j=M_j\alpha_j/2$, $M_j\in\Z$;
\item[(iii)] if $P_j$ intersects with a plane $P_0$ such that $s_0\in\Gamma$,
then $P_j\perp P_0$ and the intersection $L_0=P_j\cap P_0$ satisfies either
$L_0=\sigma L_j$ or $L_0=\sigma L_{j+1}$ for some $\sigma\in\Gamma$. The angle between $L_j$
and $L_0$ is $k\alpha_j$ with an integer $k$.
\item[(iv)] the left and right subgroups ${\bf L}$ and ${\bf R}$ in the expression
$\Gamma=({\bf L}\rl{\bf L}_K;{\bf R}\rl{\bf R}_K)$ satisfy
$\D_2\subset{\bf L}$ and $\D_2\subset{\bf R}$.
\end{itemize}
\end{lemma}

\proof
As it is noted in section \ref{sec2},  lemma \ref{cor:1}, either $\Sigma_j\cong\Z_2$ or $\Sigma_j\cong(\Z_2)^2$.
In both cases $s_j$ is an element of the group.

(i) To prove that $\alpha_j=\pi/K_j$, it is enough to remark that
$N(\Sigma_j)/\Sigma_j \cong \D_{K_j}$ (dihedral group of order $2K_j$) for some integer $K_j>1$. Then, since $s_{j+1}s_{j-1}$ is a rotation acting in $P_j$, it has to be in $D_{K_j}$, so the only possibility is that $2\alpha_j=2\pi/K_j$.

(ii) $\alpha_j=\pi/K_j$ implies that $(s_{j+1}s_{j-1})^{K_j}\xi_j=\xi_j$, therefore
$(s_{j+1}s_{j-1})^{K_j}\in\Sigma_j$. This transformation acts on
$P_j^{\perp}$ as a rotation by $2\beta_jK_j$. Since $\Sigma_j\cong\Z_2$ or
$\Sigma_j\cong(\Z_2)^2$, $2\beta_jK_j=k\pi$, which implies $\beta_j=k\alpha_j/2$.

(iii) Note that $L_0$ is one of the axes of symmetries, otherwise for some
$\rho\in \D_{K_j}$ the axis $\rho L_0$ intersects with $\kappa_j$.
Since $L_0$ is an axis of symmetry, $L_0=\sigma L_j$ or
$L_0=\sigma L_{j+1}$, and the definition of simple cycles implies that
the intersection is orthogonal.

(iv) We choose a basis in $\R^4$ such that
$\xi_2=(0,a,0,0)$ and invariant planes containing the trajectories that join
$\xi_2$ with $\xi_1$ and $\xi_3$ are
\begin{equation}\label{invp}
P_1=<{\bf e}_1,{\bf e}_2>,\ P_2=<{\bf e}_2,{\bf e}_3>.
\end{equation}
Denote by $({\bf l}_j;{\bf r}_j)$ a preimage of $s_j$ under the
homomorphism $\Phi$. We have
\begin{equation}\label{phi1}
\Phi^{-1}s_1=({\bf l}_1;{\bf r}_1)=((0,1,0,0);(0,1,0,0)),\
\Phi^{-1}s_2=({\bf l}_2;{\bf r}_2)=((0,0,0,1);(0,0,0,-1)).
\end{equation}
The group generated by ${\bf l}_1$ and ${\bf l}_2$ is $\D_2$,
and so is the one generated by  ${\bf r}_1$ and ${\bf r}_2$.
\qed

\begin{lemma}\label{lem77}
Suppose that a finite group $\Gamma^*\subset$\,O(4), $\Gamma^*\not\subset$SO(4), admits
simple heteroclinic cycles. Then the group $\Gamma=\Gamma^*\cap$\,SO(4) admits simple heteroclinic cycles.
\end{lemma}

\proof
Let $\Sigma_j^*,\Delta_j^*\subset\Gamma^*$, $j=1,\dots, m^*$, and
$\gamma^*\in\Sigma_j^*$, be the sequences of isotropy subgroups, and
the symmetry, satisfying {\bf C1}-{\bf C5} in the statement of lemma
\ref{lem12}. Define the subgroups $\Sigma_j,\Delta_j\subset\Gamma$ as
follows:
\begin{itemize}
\item If $\Sigma_j^*\cong\Z_2$ (this is satisfied or not satisfied
simultaneously for all $j$), then $\Sigma_j^*\subset$\,SO(4) and we set
$\Sigma_j=\Sigma_j^*$ and $\Delta_j=\Delta_j^*$, $j=1,m^*$.
\item If $\Sigma_j^*\cong(\Z_2)^2$, then there exists a plane reflection
$\sigma_j\in\Sigma_j^*$, $\sigma_j\in$\,SO(4). We set $\Sigma_j=<\sigma_j>$ and
$\Delta_j=<\sigma_{j-1},\sigma_j>$, $j=1,m^*$.
\item If $\gamma^*\in$\,SO(4), then $\gamma=\gamma^*$ and $m=m^*$.
\item If $\gamma^*\notin$\,SO(4), then $\gamma=(\gamma^*)^2$, $m=2m^*$,
$\Sigma_{j+m^*}=\gamma^*\Sigma_j(\gamma^*)^{-1}$ and
$\Delta_{j+m^*}=\gamma^*\Delta_j(\gamma^*)^{-1}$, $j=1,m^*$.
\end{itemize}
Evidently, $\Sigma_j,\Delta_j\subset\Gamma$, $j=1,\dots, m$, and
$\gamma\in\Sigma_j$, satisfy {\bf C1}-{\bf C5}. Hence, if the group $\Gamma^*$
admits simple heteroclinic cycles, then so does $\Gamma$.
\qed

\begin{lemma}\label{lem7}
Let $r,\ s,\ k_1,\ k_2$, $n_1$, $n_2$ and $n_3$ be integers satisfying the relation
\begin{equation}\label{eqlem7}
{n_1\over k_1}+{n_3\over rk_1}={n_2\over k_2}+{sn_3\over rk_2}=\nu.
\end{equation}
\begin{itemize}
\item[(i)] If
\begin{equation}\label{stlem7}
\hbox{$k_1$ and $k_2$ are co-prime; $r$ and $k_2-sk_1$ are co-prime}
\end{equation}
then $\nu\in \mathbb{Z}$.
\item[(ii)] If $\nu\notin \mathbb{Z}$ then at least one of the conditions in
(\ref{stlem7}) is not satisfied.
\end{itemize}
\end{lemma}

\proof
First, we notice that if $k_1=mK_1$ and $k_2=mK_2$ with $m>1$, then
$n_1=K_1$, $n_2=K_2$ and $n_3=0$ is a solution to (\ref{eqlem7}) with $\nu\notin\mathbb{Z}$. Now we suppose $k_1\wedge k_2=1$.
Assume, that there exists a solution to (\ref{eqlem7}) such that $\nu\notin\mathbb{Z}$. Since $k_1$ and $k_2$ are co-prime, for
this solution $n_3\ne rK_3$. Re-write (\ref{eqlem7}) as
$$n_1k_2-n_2k_1=n_3{sk_1-k_2\over r}.$$
If $sk_1-k_2$ and $r$ are co-prime, then the above equation does not have solutions
with $n_3\ne rK_3$; and if they are not co-prime, then it does.
\qed

\subsection{Proof of Theorems \ref{th1} and \ref{th2}}
\label{mainproofs}

\subsubsection{Proof of Theorem \ref{th1}}

According to lemma \ref{lem5}(iv), if a $\Gamma$-equivariant system possesses
a heteroclinic cycle then the left and the right groups of
$\Gamma=({\bf L}\rl{\bf L}_K;{\bf R}\rl{\bf R}_K)$ satisfy
$\D_2\subset{\bf L}$ and $\D_2\subset{\bf R}$. Such subgroups of SO(4) are
the groups 10-32 and 34-39 listed in table \ref{listSO4}. By definition of
simple heteroclinic cycles and thanks to Lemma \ref{lem5}(iii), an
admissible group $\Gamma\subset$\,SO(4) involves at least two plane reflections
$s_1$ and $s_2$, such that
\begin{itemize}
\item[I] $\dim P_1\cap P_2=1$, where $P_j=\Fix(s_j)$, $j=1,2$;
\item[II] if $P_1$ or $P_2$ intersects with a plane $P_0=\Fix(s_0)$, where
$s_0\in\Gamma$, then $P_j\perp P_0$;
\item[III] if $L'=\Fix(\Delta')$ for some $\Delta'\subset\Gamma$ satisfies
$\dim L'=1$ and $L'\subset P_j$, $j=1$ or 2, then $\Delta'\cong(\Z_2)^2$.
\end{itemize}

To study whether $\Gamma\subset$SO(4) admits simple heteroclinic cycles,
we proceed in three steps.

\medskip
In step [i] we identify all plane reflections, which are elements of the groups
10-32 and 34-39 in Table \ref{listSO4}. A plane reflection
$g=({\bf l};{\bf r})\in\Gamma=({\bf L}\rl{\bf L}_K;{\bf R}\rl{\bf R}_K)$, satisfies
\begin{equation}\label{condrl}
{\bf l}^2=(-1,0,0,0)\hbox{ and }{\bf r}^2=(-1,0,0,0).
\end{equation}
Using (\ref{finsg}) and the correspondence between ${\bf L}$
and ${\bf R}$ discussed in section \ref{sec:quaternions}, we obtain all such
pairs $({\bf l},{\bf r})$.
The results are listed in annex \ref{planereflections}.
In particular, we identify subgroups of SO(4) which do not possess plane
reflections satisfying I and II.

In step [ii] we determine the conjugacy classes of subgroups of $\Gamma$,
isomorphic to $\Z_2$,
which are generated by a plane reflection, and conjugacy classes of
$\Delta'\cong(\Z_2)^2$, such that $\Delta'$ is generated by two plane
reflections and $\dim\Fix(\Delta')=1$.
These subgroups are listed in annex \ref{conjugacyclasses} for
all $\Gamma$'s satisfying I and II.
Using lemma \ref{lemn6} we then identify those $\Gamma$'s, which do not have plane
reflections satisfying I-III.

Finally, in step [iii],
using the list in annex \ref{conjugacyclasses}
we identify those sequences $\Sigma_j$ and $\Delta_j$ which satisfy
{\bf C1}-{\bf C5} and calculate structure angles $\alpha_j$ and $\beta_j$.
They are presented in annex \ref{Sigmaj-Deltaj}.
If ${\bf l}=(\cos\omega,{\bf v}\sin\omega)$ and
${\bf r}=(\cos\omega',{\bf v}'\sin\omega')$, then the transformation
${\bf q}\to{\bf lqr}^{-1}$ is a rotation of angles $\omega\pm\omega'$ in a
pair of absolutely perpendicular planes. Let $\Sigma_j\cong\Z_2$
be represented as $\Sigma_j=\{e,\sigma_j\}$. If $\alpha_j$ and $\beta_j$
are the structure angles of heteroclinic cycles according to definition
\ref{defsa}, then the product $\sigma_{j+1}\sigma_{j-1}$ acts as rotation by angles
$2\alpha_j$ in $P_j$ and $2\beta_j$ in $P_j^{\perp}$, which allows us
to calculate the angles $\alpha_j$ and $\beta_j$ from $\sigma_{j+1}$ and
$\sigma_{j-1}$. The angle $\alpha_j$ can be also found as $\alpha_j=\pm\pi/k_j$,
where $\D_{k_j}=N(\Sigma_j)/\Sigma_j$. To find the structure angles we first
determine $\alpha_j$. Then we represent in the quaternionic form
$\sigma_{j+1}\sigma_{j-1}=
((\cos\omega,{\bf v}\sin\omega);(\cos\omega',{\bf v}'\sin\omega'))$ and note
that $2\alpha_j=\omega\pm\omega'$ and $2\beta_j=\omega\mp\omega'$, which allows
to find $\beta_j$.

\bigskip
Below we show that the groups
$$(\D_{2K_1}\rl\D_{2K_1};\D_{2K_2}\rl\D_{2K_2})\hbox{ and }
(\T\rl\Z_2;\T\rl\Z_2)$$
admit simple heteroclinic cycles, while the groups
$$(\T\rl\T;\T\rl\T),\ (\mO\rl\mO;\mO\rl\mO)\hbox{ and }
(\mO\rl\Z_1;\mO\rl\Z_1)$$
do not.  We derive the conditions (the relations between $n$, $k$, $r$
and $s$ for the first group, the restrictions on $K$ for the second) for
the groups
$$(\D_{nr}\rl\Z_{2n};\D_{kr}\rl\Z_{2k})_s\hbox{ and }
(\D_{2K}\rl\D_K;\mO\rl\T)$$
to admit simple heteroclinic cycles. For other groups the proofs are
similar and we omit them. The proofs follow from annexes
\ref{planereflections}-\ref{Sigmaj-Deltaj}
, where\\
$\bullet$ In annex \ref{planereflections} we list all plane reflections, which are elements of the groups
10-32 and 34-39. Subgroups of SO(4), which do not
possess plane reflections satisfying I and II, can be found from this list.\\
$\bullet$ In annex \ref{conjugacyclasses} for all $\Gamma$'s satisfying I and II,
we list conjugacy classes of subgroups of $\Gamma$, isomorphic to
$\Z_2$, which are generated by a plane reflection, and conjugacy classes of
$\Delta'\cong(\Z_2)^2$, $\dim\Fix(\Delta')=1$.\\
$\bullet$ In annex \ref{Sigmaj-Deltaj} we list sequences $\Sigma_j$ and $\Delta_j$ which satisfy
{\bf C1}-{\bf C5} and structure angles $\alpha_j$ and $\beta_j$.

\bigskip
{\bf The group $(\D_{2K_1}\rl\D_{2K_1};\D_{2K_2}\rl\D_{2K_2})$.}
\begin{itemize}
\item [[i]] The group $\D_n$ (see (\ref{finsg})\,) is comprised of the elements
\begin{equation}\label{dn}
\rho_n(t)=(\cos t\pi/n,0,0,\sin t\pi/n),\
\sigma_n(t)=(0,\cos t\pi/n,\sin t\pi/n,0),\ 0\le t<2n.
\end{equation}
The pairs $({\bf l};{\bf r})\in(\D_{2K_1}\rl\D_{2K_1};\D_{2K_2}\rl\D_{2K_2})$
satisfy ${\bf l}\in\D_{2K_1}$, ${\bf r}\in\D_{2K_2}$, where all possible
combinations are elements of the group. Hence, in view of (\ref{condrl}), the
plane reflections are
\begin{equation}\label{pref1}
\begin{array}{l}
\kappa_1(\pm)=((0,0,0,1);(0,0,0,\pm1)),\\
\kappa_2(n_1)=((0,\cos(n_1\theta_1),\sin(n_1\theta_1),0);(0,0,0,1)),\\
\kappa_3(n_2)=((0,0,0,1);(0,\cos(n_2\theta_2),\sin(n_2\theta_2),0)),\\
\kappa_4(n_1,n_2)=((0,\cos(n_1\theta_1),\sin(n_1\theta_1),0);
(0,\cos(n_2\theta_2),\sin(n_2\theta_2),0)),
\end{array}\end{equation}
where $\theta_1=\pi/(2K_1)$, $\theta_2=\pi/(2K_2)$, $0\le n_1<4K_1$ and
$0\le n_2<4K_2$. Lemma \ref{lemn6} implies that for any $n_1$ and $n_2$ the plane reflections
$s_1=\kappa_2(n_1)$ and $s_2=\kappa_3(n_2)$ satisfy I and II.

\item [[ii]] In the group $\D_n$ the elements
$(0,\cos(t\pi/n),\sin(t\pi/n),0)$ split into two conjugacy classes,
corresponding to odd and even $t$. The elements $(0,0,0,1)$ and $(0,0,0,-1)$ are
conjugate. Therefore, the group has nine conjugacy classes of isotropy
subgroups $\Sigma\cong\Z_2$:
\begin{equation}\label{pref11}
\begin{array}{l}
\{e,\kappa_1(\pm)\},\\
\{e,\kappa_2(n_1)\}:\ n_1\hbox{ even or odd},\\
\{e,\kappa_3(n_2)\}:\ n_2\hbox{ even or odd},\\
\{e,\kappa_4(n_1,n_2)\}:\ n_1\hbox{ even or odd, }n_2\hbox{ even or odd}.\\
\end{array}\end{equation}
For a subgroup of SO(4), a symmetry axis is an intersection of symmetry
planes. Any plane $\Fix(\kappa_2(n_1))$ intersects with any
$\Fix(\kappa_3(n_2))$ and the line of intersection also belongs to the
plane $\Fix(\kappa_4(n_1-K_1,n_2+K_2))$.
The isotropy subgroup of the line is
\begin{equation}\label{pref12}
\Delta=\{e,\kappa_2(n_1),\kappa_3(n_2),\kappa_4(n_1-K_1,n_2+K_2)\}.
\end{equation}
The isotropy subgroups (\ref{pref12}) split into four conjugacy
classes, corresponding to odd and even $n_1$ and $n_2$.
Note that $\Delta\cong(\Z_2)^2$ and the planes $\Fix(\kappa_2(n_1))$ and
$\Fix(\kappa_3(n_2))$ do not have other symmetry axes. Therefore, $s_1$ and
$s_2$ satisfy III.

\item [[iii]] We set:
\begin{equation}\label{pref13}
\begin{array}{l}
\Sigma_1=\{e,\kappa_2(0)\},\ \Sigma_2=\{e,\kappa_3(1)\},\
\Sigma_3=\{e,\kappa_2(1)\},\ \Sigma_4=\{e,\kappa_3(0)\},\\
\Delta_1=\{e,\kappa_2(0),\kappa_3(0),\kappa_4(-K_1,K_2)\},\
\Delta_2=\{e,\kappa_2(0),\kappa_3(1),\kappa_4(-K_1,K_2+1)\},\\
\Delta_3=\{e,\kappa_2(1),\kappa_3(1),\kappa_4(-K_1+1,K_2+1)\},\
\Delta_4=\{e,\kappa_2(1),\kappa_3(0),\kappa_4(-K_1+1,K_2)\}.
\end{array}\end{equation}
By construction, the sequences $\Sigma_j,\Delta_j,j=1,\ldots,4$, with
$\gamma=e$ satisfy conditions {\bf C1}-{\bf C5} of lemma \ref{lem12}.

Since $N(\Sigma_1)/\Sigma_1=\D_{4K_2}$, we have
$\alpha_1=\pm\pi/(4K_2)=\pm\theta_2/2$. To find $\beta_1$, we calculate
$$\sigma_2\sigma_4=\kappa_3(1)\kappa_3(0)=
((-1,0,0,0);(-\cos(n_2\theta_2),0,0,\sin(n_2\theta_2))),$$
which implies that
$$2\alpha_1=\pi\pm(\pi+\theta_2)\hbox{ and }2\beta_1=\pi\mp(\pi+\theta_2).$$
Hence, $\alpha_1=-\theta_2/2$ and $\beta_1=\pi-\alpha_1$. Similarly, we calculate
that $\alpha_3=\alpha_1$, $\alpha_2=\alpha_4=-\theta_1/2$ and
$\beta_j=\pi-\alpha_j$, $j=$2,3,4.
\end{itemize}

In fact, the group has two more isotropy types of symmetry axes, which are
the intersections of $\kappa_1(\pm)$ with $\kappa_4(n_1,n_2)$. However,
the isotropy groups $\Delta$ of these axes satisfy $\Delta\cong(\Z_2)^2$ only
if $K_1$ and $K_2$ are co-prime. The goal of the study is to prove
existence of heteroclinic cycles, and not to find the largest heteroclinic network
which can possibly exist in a $\Gamma$-equivariant system.
Since the isotropy subgroups (\ref{pref13}) satisfy {\bf C1}-{\bf C5}, the
additional axes are not discussed. Similarly, we do not discuss the largest possible heteroclinic
networks admitted by other groups.

\bigskip
{\bf The group $(\D_{nr}\rl\Z_{2n};\D_{kr}\rl\Z_{2k})_s$.}
\begin{itemize}
\item [[i]] The condition
$\D_2\subset\D_{nr}$,$\D_{kr}$, implies that $\Gamma$ is either \\
$(\D_{2K_1r}\rl\Z_{4K_1};\D_{2K_2r}\rl\Z_{4K_2})_s$ with odd $r$, or
$(\D_{2K_1r}\rl\Z_{2K_1};\D_{2K_2r}\rl\Z_{2K_2})_s$. Because of (\ref{condrl}) and
(\ref{dn}), the reflections in the group
$(\D_{2K_1r}\rl\Z_{4K_1};\D_{2K_2r}\rl\Z_{4K_2})_s$ are
\begin{equation}\label{pref2}
\begin{array}{l}
\kappa_1(\pm)=((0,0,0,1);(0,0,0,\pm1)),\\
\kappa_2(n_1,n_2,n_3)=((0,\cos(n_1\theta_1+n_3\theta_1^*),
\sin(n_1\theta_1+n_3\theta_1^*),0); \\ ~~~~~~~~~~~~~~~~~~~~~~~~~~~~~~~~~~~~~~
(0,\cos(n_2\theta_2+n_3s\theta_2^*),\cos(n_2\theta_2+n_3s\theta_2^*),0)),
\end{array}\end{equation}
where $0\le n_j<4K_j$, $\theta_j=\pi/(2K_j)$ and $\theta_j^*=\theta_j/r$,
$j=1,2$, $0\le n_3<r$. Denote by $P(n_1,n_2,n_3)$ the fixed-point subspace of
$\kappa_2(n_1,n_2,n_3)$. Lemma \ref{lemn6} implies that planes $P(n_1,n_2,n_3)$
and $P(n_1',n_2',n_3')$ intersect if
\begin{equation}\label{eqvi}
\cos((n_1-n_1')\theta_1+(n_3-n_3')\theta_1^*)=
\cos((n_2-n_2')\theta_2+(n_3-n_3')s\theta_2^*).
\end{equation}
By lemma \ref{lem7}, if
\begin{equation}\label{condno1}
\hbox{$K_1$ and $K_2$ are co-prime, $r$ and $K_2-sK_1$ are co-prime,}
\end{equation}
then the only solutions to this equation are
$(n_1-n_1')\theta_1+(n_3-n_3')\theta_1^*=M_1\pi/2$,
$(n_2-n_2')\theta_2+(n_3-n_3')s\theta_2^*=M_2\pi/2$, i.e. any intersection is
orthogonal. If (\ref{condno1}) is not satisfied, then there exist solutions to (\ref{eqvi})
with $\cos((n_1-n_1')\theta_1+(n_3-n_3')\theta_1^*)\ne 0,\pm1$, hence the
intersection is non-orthogonal. If (\ref{condno1}) holds true, then
$s_1=\kappa_2(0,0,0)$ and $s_2=\kappa_2(K_1,K_2,0)$ satisfy I and II.

The elements of the group
$(\D_{2K_1r}\rl\Z_{2K_1};\D_{2K_2r}\rl\Z_{2K_2})_s$
which are plane reflections are different for odd and even $K_1$ and $K_2$
(note that the case when they are both even was considered above). If $K_1$ is
even and $K_2$ is odd, then plane reflections are given by
\begin{equation}\label{pref3}
\begin{array}{ll}
\kappa_1(n_1,n_2,n_3)=&((0,\cos(2n_1\theta_1+n_3\theta_1^*),
\sin(2n_1\theta_1+n_3\theta_1^*),0);\\
&(0,\cos(2n_2\theta_2+n_3s\theta_2^*),\cos(2n_2\theta_2+n_3s\theta_2^*),0)),\\
\kappa_2(n_1,n_2,n_3)=&((0,\cos((2n_1+1)\theta_1+n_3\theta_1^*),
\sin((2n_1+1)\theta_1+n_3\theta_1^*),0);\\
&(0,\cos((2n_2+1)\theta_2+n_3s\theta_2^*),\cos((2n_2+1)\theta_2+n_3s\theta_2^*),0)),
\end{array}\end{equation}
where $0\le n_j<2K_j$, $j=1,2$. It can be easily shown that whenever
$\kappa_i(n_1,n_2,n_3)$ and $\kappa_j(n_1',n_2',n_3')$, $i,j=1,2$, intersect,
the intersection is non-orthogonal.

The group $\Gamma=(\D_{2K_1r}\rl\Z_{2K_1};\D_{2K_2r}\rl\Z_{2K_2})_s$,
where $K_1$ and $K_2$ odd, involves plane reflections
\begin{equation}\label{pref4}
\begin{array}{ll}
\kappa_1(\pm)=&((0,0,0,1);(0,0,0,\pm1)),\\
\kappa_2(n_1,n_2,n_3)=&((0,\cos(2n_1\theta_1+n_3\theta_1^*),
\sin(2n_1\theta_1+n_3\theta_1^*),0);\\
&(0,\cos(2n_2\theta_2+n_3s\theta_2^*),\cos(2n_2\theta_2+n_3s\theta_2^*),0)),\\
\kappa_3(n_1,n_2,n_3)=&((0,\cos((2n_1+1)\theta_1+n_3\theta_1^*),
\sin((2n_1+1)\theta_1+n_3\theta_1^*),0);\\
&(0,\cos((2n_2+1)\theta_2+n_3s\theta_2^*),\cos((2n_2+1)\theta_2+n_3s\theta_2^*),0)).
\end{array}\end{equation}
Lemma \ref{lem7} implies that whenever
\begin{equation}\label{condno2}
\hbox{$K_1$ and $K_2$ are co-prime, $r$ and $(K_2\pm sK_1)/2$ are co-prime,}
\end{equation}
a plane fixed by
$\kappa_j(n_1,n_2,n_3)$, $j=1$ or 2, intersect only orthogonally with another plane
fixed by a plane reflection. Hence, we set
$s_1=\kappa_2(0,0,0)$ and $s_2=\kappa_2((K_1-1)/2,(K_2-1)/2,0)$.

For the group $\Gamma=(\D_{2K_1r}\rl\Z_{K_1};\D_{2K_2r}\rl\Z_{K_2})_s$, where
$K_1$ and $K_2$ are odd, by lemma \ref{lem7} the planes fixed by elements of the group intersect
only orthogonally if and only if
\begin{equation}\label{condno3}
\hbox{$K_1$ and $K_2$ are co-prime, $r$ and $(K_2\pm sK_1)/2$ are co-prime,
$r$ and $(K_2\pm sK_1)/4$ are co-prime,}
\end{equation}
where plus or minus are taken so that the ratios are integer.

\item [[ii]] In the group $\D_{2n}$ (see (\ref{dn})\,)
the elements $\rho_{2n}(n)=(0,0,0,1)$ and $\rho_{2n}(3n)=(0,0,0,-1)$ are
conjugate by $\sigma_{2n}(t)$. The group
$(\D_{nr}\rl\Z_{2n};\D_{kr}\rl\Z_{2k})_s$ involves $\sigma$'s only in pairs
$(\sigma_{nr}(t_1);\sigma_{kr}(t_2))$, therefore $\kappa_1(+)$ and
$\kappa_1(-)$ are not conjugate in this group. The splitting of $\kappa_2$ and
$\kappa_3$ into conjugacy classes depends on whether $K_1$, $K_2$ and $r$
are even or odd. Here we consider only the case of
$(\D_{2K_1r}\rl\Z_{4K_1};\D_{2K_2r}\rl\Z_{4K_2})_s$, where $K_1$, $K_2$, $r$
and $s$ satisfy (\ref{condno1}), $K_1$, $K_2$ and $r$ are odd. The cases
of other parities are similar and we do not present them. Under this assumption,
the reflections $\kappa_2(n_1,n_2,n_3)$ in (\ref{pref2}) split into
four conjugacy classes, a class is categorised by whether the sums $n_1+n_3$ and
$n_2+n_3$ are even or odd. By arguments presented in the part [i], if
(\ref{condno1}) is satisfied, then the reflections $s_1$ and $s_2$ satisfy III.

\item [iii] We set:
\begin{equation}\label{pref23}
\begin{array}{l}
\Sigma_1=\{e,\kappa_2(0,0,0)\},\ \Sigma_2=\{e,\kappa_2(K_1,K_2,0)\},\\
\Delta_1=\{e,\kappa_1(-),\kappa_2(0,0,0),\kappa_2(K_1,3K_2,0)\},\\
\Delta_2=\{e,\kappa_1(+),\kappa_2(0,0,0),\kappa_2(K_1,K_2,0)\},
\end{array}\end{equation}
and
$\gamma=((1,0,0,0);(0,0,0,1))$. Since
$\gamma\kappa_2(n_1,n_2,n_3)\gamma^{-1}=\kappa_2(n_1+2K_1,n_2+2K_2,n_3)$,
the sequences $\Sigma_j,\Delta_j,j=1,\ldots,2$,
satisfy conditions {\bf C1}-{\bf C5} of lemma \ref{lem12}.

We have $N(\Sigma_1)/\Sigma_1=\D_2$ and
$\sigma_{-1}\sigma_2=\gamma\sigma_2\gamma^{-1}\sigma_2=((-1,0,0,0);(1,0,0,0))$,
therefore $\alpha_1=\pi/2$ and $\beta_1=\pi/2$. Similarly,
$\alpha_2=\pi/2$ and $\beta_2=\pi/2$.
\end{itemize}

\bigskip
{\bf The group} $(\T\rl\T;\T\rl\T)$.
\begin{itemize}
\item [[i]] The pairs $({\bf l};{\bf r})\in(\T\rl\T;\T\rl\T)$ satisfy
${\bf l}\in\T$, ${\bf r}\in\T$, where all possible combinations are
elements of the group. Hence, the plane reflections are
$$\kappa(\pm,r,s)=\pm(\rho^r{\bf u};\rho^s{\bf u}),$$
where ${\bf u}=(0,0,0,1)$ and the permutation $\rho$ acts as
$\rho(a,b,c,d)=(a,c,d,b)$. By lemma \ref{lemn6}, whenever two $\kappa$'s
intersect, the intersection is orthogonal. Taking $s_1=\kappa(+,0,0)$ and
$s_2=\kappa(+,1,1)$, we get plane reflections satisfying I and II.
\item [[ii]] The quaternions $(0,0,1,0)$
and $(0,1,0,0)$ are conjugate by $(a,b,b,a)$, the quaternions $(0,0,1,0)$ and
$(0,0,-1,0)$ are conjugate by $(0,a,b,0)$, hence all plane reflections
in the group are conjugate. The plane fixed by $\kappa(\pm,r,s)$ intersects
with the ones $\kappa(\pm,r+t,s+t')$, $t,t'=1,2$, the lines of intersection can
be of two isotropy type, involving the following plane reflections:
\begin{equation}\label{iiT}
\begin{array}{l}
(a):\ \kappa(\pm,r,s),\ \kappa(\pm,r+1,s+1),\ \kappa(\pm,r+2,s+2);\\
(b):\ \kappa(\pm,r,s),\ \kappa(\pm,r+1,s+2),\ \kappa(\pm,r+2,s+1).
\end{array}\end{equation}
In both cases there exists a symmetry $\sigma\in(\T\rl\T;\T\rl\T)$ which cyclically
interchanges the three planes fixed by $\kappa(\pm,r+j,s+k)$, $j,k=0,1,2$.
Hence, the group $(\T\rl\T;\T\rl\T)$ does not satisfy III.
\end{itemize}

\medskip
{\bf The group} $(\T\rl\Z_2;\T\rl\Z_2)$.
\begin{itemize}
\item [[i]] Elements of the group are the pairs
$({\bf l};{\bf r})$ such that ${\bf l}\in\T$ and ${\bf r}=\pm{\bf l}$.
Therefore, the group involves plane reflections
$$\kappa(\pm,r)=\pm(\rho^r{\bf u};\rho^r{\bf u}).$$
The planes $\Fix(\kappa(\pm,r))$ and $\Fix(\kappa(\pm,s))$ intersect whenever
$r\ne s$ and the intersection is orthogonal. The plane reflections
$s_1=\kappa(-,0)$ and $s_2=\kappa(-,1)$ satisfy I and II.
\item [[ii]] The plane reflections split into two
conjugacy classes: $\kappa(+,r)$ and $\kappa(-,r)$. There are two isotropy
types of one-dimensional subspaces, their symmetry groups involve
the following plane reflections:
\begin{equation}\label{iiTZ}
\begin{array}{l}
(a):\ \kappa(+,r),\ \kappa(+,r+1),\ \kappa(+,r+2);\\
(b):\ \kappa(+,r),\ \kappa(-,r+1),\ \kappa(-,r+2).
\end{array}\end{equation}
In the former case the three plane reflections are cyclically conjugate by a
symmetry\\
$((1/2,1/2,1/2,1/2);(1/2,1/2,1/2,1/2))$ in $(\T\rl\Z_2;\T\rl\Z_2)$, hence
the isotropy subgroup of this line is $\D_3$. In the latter case the isotropy
subgroup is $(\Z_2)^2$.
\item [[iii]] Setting
$$\Sigma_1=\{e,\kappa(-,0)\},\
\Delta_1=\{e,\kappa(+,2),\kappa(-,0),\kappa(-,1)\}$$
and $\gamma=((1/2,1/2,1/2,1/2);(1/2,1/2,1/2,1/2))$,
we get the sequences (with $m=1$) satisfying {\bf C1}-{\bf C5}.
The structure angles are $\alpha_1=\pi/2$ and $\beta_1=\pi$.
\end{itemize}

\bigskip
{\bf The group} $(\mO\rl\mO;\mO\rl\mO)$.
\begin{itemize}
\item [[i]] The pairs $({\bf l};{\bf r})\in(\mO\rl\mO;\mO\rl\mO)$ are any combinations of
${\bf l}\in\mO$ and ${\bf r}\in\mO$. Hence, the plane reflections are
$$\kappa_1(\pm,r,s)=\pm(\rho^r{\bf u};\rho^s{\bf u}),\
\kappa_2(\pm,r,s,\pm)=\pm(\rho^r{\bf u};\rho^s{\bf v}_{\pm}),$$
$$\kappa_3(\pm,r,\pm,s)=(\rho^r{\bf v}_{\pm};\rho^s{\bf u}),\
\kappa_4(\pm,r,\pm,s,\pm)=(\rho^r{\bf v}_{\pm};\rho^s{\bf v}_{\pm}),$$
where ${\bf u}=(0,0,0,1)$, ${\bf v}_{\pm}=(0,1,\pm1,0)/\sqrt{2}$ and
the permutation $\rho$ acts as $\rho(a,b,c,d)=(a,c,d,b)$. Planes
fixed by $\kappa_1$ and $\kappa_4$ intersect non-orthogonally and so do the ones
fixed by $\kappa_2$ and $\kappa_3$. Therefore, the group does not have
plane reflections satisfying I and II.
\end{itemize}

\bigskip
{\bf The group} $(\mO\rl\Z_1;\mO\rl\Z_1)$.
\begin{itemize}
\item [[i]] The group is comprised of the pairs $({\bf l};{\bf r})$, such that
${\bf l}\in\mO$ and ${\bf l}={\bf r}$. The plane reflections are
$$\kappa_1(r)=(\rho^r{\bf u};\rho^r{\bf u})\hbox{ and }
\kappa_2(r,\pm)=(\rho^r{\bf v}_{\pm};\rho^r{\bf v}_{\pm}).$$
Since the planes fixed by $\kappa_1$ and $\kappa_2$ intersect non-orthogonally,
the group does not admit heteroclinic cycles.
\end{itemize}

\bigskip
{\bf The group} $\Gamma=(\D_{2K}\rl\D_K;\mO\rl\T)$, $K$ even.
\begin{itemize}
\item [[i]] The group $(\D_{2K}\rl\D_K;\mO\rl\T)$ is comprised of the pairs
$({\bf l};{\bf r})$, where either\\
${\bf l}\in\D_K$ and ${\bf r}\in\T$, or\\
${\bf l}\in\D_{2K}\setminus\D_K$ and ${\bf r}\in\mO\setminus\T$.\\
Therefore, for even $K$ the group has the following plane reflections:
\begin{equation}\label{iDO}
\begin{array}{l}
\kappa_1(\pm,r)=((0,0,0,\pm1);\rho^r{\bf u}),\\
\kappa_2(n,r)=((0,\cos(2n\theta),\sin(2n\theta),0);\rho^r{\bf u}),\\
\kappa_3(n,r,\pm)=((0,\cos((2n+1)\theta),\sin((2n+1)\theta),0);
\rho^r{\bf v}_{\pm}),
\end{array}\end{equation}
where $\theta=\pi/(2K)$ and $0\le n\le 2K$.
By lemma \ref{lemn6}, if $K=2(2k+1)$ then the planes fixed by
$\kappa_2$ and $\kappa_3$ intersect non-orthogonally. Otherwise, plane
reflections $s_1=\kappa_2(0,0)$ and $s_2=\kappa_2(K/2,1)$ satisfy I and II.
\item [[ii]] The group has three conjugacy classes of isotropy subgroups
satisfying $\dim\Fix(\Sigma))=2$, they are
\begin{equation}\label{iiDO}
\{e,\kappa_1(\pm,r)\},\ \{e,\kappa_2(n,r)\},\ \{e,\kappa_3(n,r,\pm)\}.
\end{equation}
For $K\ne2(2k+1)$ it has two isotropy types of symmetry axes, one of
which has the isotropy subgroup
\begin{equation}\label{iiaDO}
\{e,\kappa_1(\pm,r),\kappa_2(n,r+1),\kappa_2(n+K/2,r+2)\},
\end{equation}
isomorphic to $(\Z_2)^2$. (The other axis has isotropy subgroup
generated by two $\kappa_3$, it can be isomorphic to $(\Z_2)^2$, or it can
be not, depending on $K$.) The planes fixed by $\kappa_2$ contains
only symmetry axes with the group (\ref{iiaDO}). Therefore, III holds true.
\item [[iii]] The isotropy subgroups
$$\Sigma_1=\{e,\kappa_2(0,0)\},\
\Delta_1=\{e,\kappa_1(+,1),\kappa_2(0,0),\kappa_2(K/2,2)\},$$
and the symmetry $\gamma=((1,0,0,1)/\sqrt{2};(1,1,1,1)/2)$
satisfy conditions {\bf C1}-{\bf C5} with $m=1$. The structure angles of
this homoclinic cycle are $\alpha_1=\pi/4$ and $\alpha_2=\pi/4$
\end{itemize}
\qed

\subsubsection{Proof of theorem \ref{th2}}

Recall, that a group $\Gamma^*\in$\,O(4), $\Gamma^*\notin$\,SO(4), can be
decomposed as
$$\Gamma^*=\Gamma\oplus\sigma\Gamma,\hbox{ where $\Gamma\subset$\,SO(4)
and $\sigma\notin$\,SO(4)},$$
where in the quaternion form
$\Phi^{-1}\Gamma=({\bf L}\rl{\bf L}_K;{\bf R}\rl{\bf R}_K)$, the groups ${\bf L}$
and ${\bf R}$ are isomorphic, and so are ${\bf L}_K$ and ${\bf R}_K$.
A reflection $\sigma:{\bf q}\to{\bf a}\tilde{\bf q}{\bf b}$ is written as
$\sigma=({\bf a},{\bf b})^*$. By lemma \ref{lem77}, if the group $\Gamma^*$
admits simple heteroclinic cycles, then so does $\Gamma$.

Admissible subgroups of $\Gamma\subset$\,SO(4) are listed in Theorem \ref{th1},
the ones which have isomorphic left and right groups are:
\begin{equation}\label{list3}
\renewcommand{\arraystretch}{1.5}
\begin{array}{l}
(\D_{2K}\rl\D_{2K};\D_{2K}\rl\D_{2K}),\ (\D_{2r}\rl\Z_{4};\D_{2r}\rl\Z_{4}),\
(\D_{2r}\rl\Z_{2};\D_{2r}\rl\Z_{2}),\\
(\D_{2K}\rl\D_{K};\D_{2K}\rl\D_{K}),\
(\T\rl\Z_2;\T\rl\Z_2),\ (\D_{2r}\rl\Z_1;\D_{2r}\rl\Z_1).
\end{array}\end{equation}
A reflection $\sigma\notin$\,SO(4) has $\pm1$ for two of its eigenvalues,
the other two being ${\rm e}^{\pm{\rm i}\omega}$.

\medskip
First, we consider $\omega=k\pi$. If $\omega=0$, then $\sigma$ is a reflection
about a three-dimensional hyperplane orthogonal
to a vector ${\bf e}$, leaving unchanged all vectors in the
hyperplane and reversing all orthogonal. If $\omega=\pi$, then $\sigma$ is
an axial reflection about an axis along a vector ${\bf e}'$.
Any plane $P_0$ fixed by a subgroup $\Sigma_0\subset\Gamma$ is mapped by $\sigma$
to a plane (perhaps, the same), fixed by $\Sigma_0'\subset\Gamma$.
If $P_j$ is one of the planes involved in a simple heteroclinic cycle,
then the orthogonal complement to ${\bf e}$, or to ${\bf e}'$, which we
denote by $V$, is either orthogonal to $P_j$, or $P_j\subset V$.
Since this holds true for all $1\le j\le m$,
the planes $P_j$ are coordinate planes in an appropriate basis,
structure angles are multiples of $\pi/2$ and ${\bf e}$ (or
${\bf e}'$) is a basis vector. The groups in (\ref{list3}) that
have structure angles multiples of $\pi/2$ are
$$(\D_4\rl\Z_2;\D_4\rl\Z_2),\ (\D_4\rl\Z_1;\D_4\rl\Z_1),\ (\D_2\rl\Z_2;\D_2\rl\Z_2),\
(\T\rl\Z_2;\T\rl\Z_2),\ (\D_2\rl\Z_1;\D_2\rl\Z_1).$$
For the first two group the direction of $L_1$ can be taken as
$(0,1,0,1)/\sqrt{2}$, for the next two groups as $(0,1,0,0)$ and for the last
as $(1,0,0,0)$. Hence, we obtain the first five groups listed
in the statement of theorem \ref{th2}.

\medskip
Second, we consider $\omega\ne 0,\pi$. The symmetry $\sigma$ maps any $P_j$
into another plane, which does not belong to the group orbit of $P_j$ in
$\Gamma$, because otherwise the isotropy subgroup of $L_j\subset P_j$ has
elements of order more than two. For $(\D_{2K}\rl\D_{2K};\D_{2K}\rl\D_{2K})$ the
only possibility is $\sigma:\ P_j\to P_{j+2}$, and therefore $L_1\to L_3$
and $L_2\to L_4$. Hence $\sigma^2$ maps $P_j\to P_{j+4}$ for any $j$.
For this group, there exists a heteroclinic cycle with four equilibria,
implying that $\sigma^2$ is an identity, which is possible only if $\sigma$
is an axial reflection, or a reflection about a three-dimensional hyperplane. Therefore,
there is no heteroclinic group in O(4), which has
$(\D_{2K}\rl\D_{2K};\D_{2K}\rl\D_{2K})$ as a reflection subgroup with
$\omega\ne 0,\pi$. For $(\T\rl\Z_2;\T\rl\Z_2)$ such a $\sigma$ does not exist, because
the group has only one group orbit of fixed planes.

For other groups in (\ref{list3}) the heteroclinic cycle
(see annex \ref{Sigmaj-Deltaj}
) involves
two group orbits of planes, hence $\sigma:\ P_j\to P_{j+1}$. Since for all
groups, except for $(\D_{2K}\rl\D_K;\D_{2K}\rl\D_K)$, $\alpha_2$ and $\beta_2$ are
multiples of $\pi/2$, they do not give rise to subgroups of O(4), different from
already obtained. For $(\D_{2K}\rl\D_K;\D_{2K}\rl\D_K)$ the condition
$\sigma:\ P_j\to P_{j+1}$ determines $\sigma$, up to multiplication by
some $\gamma\in\Gamma$.
\qed

\section{Examples}\label{sec4}
In this section we provide some examples of simple heteroclinic cycles of type A in $\R^4$. We will also give an example of a pseudo-simple heteroclinic cycle.

\subsection{Simple heteroclinic cycles of type A}
\subsubsection{The simplest case}
Consider the following transformations in $\R^4$:
\begin{eqnarray*}
\kappa_1:~&~(x_1,x_2,x_3,x_4)\mapsto (x_1,x_2,-x_3,-x_4) \\
\kappa_2:~&~(x_1,x_2,x_3,x_4)\mapsto (-x_1,x_2,x_3,-x_4) \\
\kappa_3:~&~(x_1,x_2,x_3,x_4)\mapsto (-x_1,-x_2,x_3,x_4)
\end{eqnarray*}
They generate a group $\Gamma_0$ which is isomorphic to $\Z_2^3$, note however the difference with the case $B^+_2$ in Theorem \ref{theorem:typesBC}. There is no invariant hyperplane, however each $\kappa_j$ has a planar fixed-point subspace and there are overall 6 such invariant planes. Moreover, each plane contains two axes of symmetry, which are the coordinate axes. In the list of Theorem 2, this group is
$(\D_{2}\rl\Z_{2};\D_{2}\rl\Z_{2})$ (the group $(\D_{nr}\rl\Z_{2n};\D_{kr}\rl\Z_{2k})_s$
with $m=n=1$ and $r=2$).
In terms of quaternionic presentation, we have
$$\kappa_1=\left[i,i\right], \kappa_2=\left[k,-k\right], \kappa_3=\left[i,-i\right]$$
where $i$, $j$, $k$ are the usual quaternion basis "imaginary" elements. \\
Remark that $-I\in\Gamma_0$ acts non-trivially in $\R^4$. Simple robust heteroclinic cycles can easily be built from the knowledge of the general equivariant smooth vector fields. Indeed, one can easily check the following lemma (using Schwarz theorem on the structure of equivariant vector fields under smooth compact group actions):
\begin{lemma}
Every smooth, $\Gamma_0$ equivariant differential system has the following form
\begin{eqnarray*}
\dot x_1 &=& a_1(x_1^2,x_2^2,x_3^2,x_4^2,\theta)x_1 + b_1(x_1^2,x_2^2,x_3^2,x_4^2,\theta)x_2x_3x_4 \\
\dot x_2 &=& a_2(x_1^2,x_2^2,x_3^2,x_4^2,\theta)x_2 + b_2(x_1^2,x_2^2,x_3^2,x_4^2,\theta)x_1x_3x_4 \\
\dot x_3 &=& a_3(x_1^2,x_2^2,x_3^2,x_4^2,\theta)x_3 + b_3(x_1^2,x_2^2,x_3^2,x_4^2,\theta)x_1x_2x_4 \\
\dot x_4 &=& a_4(x_1^2,x_2^2,x_3^2,x_4^2,\theta)x_4 + b_4(x_1^2,x_2^2,x_3^2,x_4^2,\theta)x_1x_2x_3
\end{eqnarray*}
where $\theta=x_1x_2x_3x_4$ and $a_j$, $b_j$ are smooth functions.
\end{lemma}
It is then an elementary computation to check that the conditions of existence of
a robust heteroclinic cycle connecting equilibria on the symmetry axes are generically fulfilled.

\subsubsection{A non-trivial example}
This example was studied first in the context of pattern formation on the hyperbolic plane \cite{fach}.
Let $\Gamma_1$ be the group generated by the following $4\times 4$ matrices:
 \begin{equation}\label{eq:matchi11}
\kappa=\left[ \begin{matrix}
 0 & 0 & 0 & -1\\
 0 & 0 & 1 & 0\\
 0 & 1 & 0 & 0\\
 -1 & 0 & 0 & 0
\end{matrix}
 \right],\quad \rho =\frac{\sqrt{2}}{2}\left[ \begin{matrix}
 0 & 1 & 1 & 0\\
 -1 & 0 & 0 & -1\\
 1 & 0 & 0 & -1\\
  0 & -1 & 1 &0
\end{matrix}
 \right], \quad \sigma=\frac{\sqrt{2}}{2} \left[ \begin{matrix}
 0 & 0 & -1 & 1\\
 0 & 0 & 1 & 1\\
 -1 & 1 & 0 & 0\\
  1 & 1 & 0 &0
\end{matrix}
 \right]
\end{equation}
This group has 96 elements. The generators can be identified with the following elements in the quaternionic presentation:
$$
\kappa=\left[i,j\right] , \rho=\frac{\sqrt{2}}{2}\left[1-k,i\right] , \sigma=\frac{\sqrt{2}}{2}\left[j+k,i\right]
$$
In the nomenclature of Theorem 2, $\Gamma_1=(\mO\rl\T;\D_{2}\rl\D_{1})$. \\
The following groups are 4 elements subgroups of $\Gamma_1$. They are isomorphic but belong to different conjugacy classes:
$$\widetilde C_{2\kappa}=\langle\sigma,\kappa\rangle \text{ and } \widetilde C'_{2\kappa}=\langle\rho^2\sigma\rho^{-2},\kappa\rangle.$$
The action of $\Gamma_1$ admits the following lattice of isotropy types \cite{fach}, where $\kappa'=\rho\kappa$ is not conjugated to $\kappa$.
\begin{figure}[h]
\begin{center}
$
\begin{psmatrix}[colsep=1cm]
(1) & & \widetilde C_{2\kappa} &  & \widetilde C'_{2\kappa} &  \\ (2) &
 \langle\kappa\rangle && \langle\sigma\rangle && \langle\kappa'\rangle
\end{psmatrix}
$
\end{center}
\ncline[nodesep=3pt]{1,3}{2,2}
\ncline[nodesep=3pt]{1,3}{2,4}
\ncline[nodesep=3pt]{1,3}{2,6}
\ncline[nodesep=3pt]{1,5}{2,2}
\ncline[nodesep=3pt]{1,5}{2,4}
\ncline[nodesep=3pt]{1,5}{2,6}
\end{figure}
The numbers in parentheses are the dimensions of the corresponding fixed-point subspaces.  Moreover the planes $\text{Fix}(\sigma)$ and $\text{Fix}(\kappa')$ contain one copy of each type of symmetry axes, while $\text{Fix}(\kappa)$ contains two copies of each. \\
The general form of $\Gamma_1$ equivariant vector fields is complicated but the polynomial form up to degree 5 has been computed in \cite{fach} and it was shown that a codimension 1 bifurcation from the trivial equilibria leads to robust heteroclinic cycles. These cycles are simple (as is clear from the isotropy subgroups). Also observe that there are in fact two types of cycles, hence a heteroclinic network. Their asymptotic stability depends upon terms of order 7.

\subsection{A pseudo-simple heteroclinic cycle} \label{subsec:simple}
Here we show that pseudo-simple cycles exist. An example is the (unique) four dimensional irreducible representation of the group $GL(2,3)$ ($2\times 2$ invertible matrices over the field $\Z_3$). This group is generated by the elements $\rho$ (order 8) and $\sigma$ (order 2) below:
\begin{equation*}
\rho = \left(\begin{array}{cc}0&2\\2&2\end{array}\right),~\sigma = \left(\begin{array}{cc}2&0\\0&1\end{array}\right)
\end{equation*}
The group has 8 conjugacy classes and exactly one 4-dimensional irreducible representation. Writing $\epsilon=\sigma\rho^{-1}$, the conjugacy classes and character table of this representation is given

\pagebreak
\noindent
below (see \cite{Lang}):
\begin{table}[h]
\begin{center}
\begin{tabular}{|c|c|c|c|c|c|c|c|c|} \hline representative & $Id$ & $\rho$ & $\rho^2$ & $-Id$ & $\rho^5$ & $\sigma$ & $\epsilon$ & $-\epsilon$ \\\hline order & 1 & 8 & 4 & 2 & 8 & 2 & 3 & 6 \\\hline \# elements & 1 & 6 & 6 & 1 & 6 & 12 & 8 & 8 \\\hline character & 4 & 0 & 0 & -4 & 0 & 0 & 1 & -1  \\\hline \end{tabular}
\end{center}
\end{table}

\noindent From this table and using the trace formula for the computation of the dimension of fixed-point subspaces \cite{cl2000} one finds that there are exactly two submaximal isotropy types: their group representatives are $\Sigma_1=\langle\sigma\rangle$ and $\Sigma_2=\langle\epsilon\rangle$. Their fixed-point subspaces have dimension 2. Moreover each of these planes contains exactly one copy of each of the two types of symmetry axes, the isotropy of which are isomorphic to the dihedral group $\D_3$ but are not conjugate in $GL(2,3)$. From this and using either the same proof as in lemma \ref{lem12} or by explicit computation of an equivariant vector field, one can show the existence of robust heteroclinic cycles between equilibria on the symmetry axes. Clearly these equilibria have isotropies which fall into cases 2 or 3 of lemma \ref{prop:isotypic decomp}: $\Sigma_2\cong\Z_3$ and $\Delta_2\cong \D_3$,
which implies that the heteroclinic cycles are pseudo-simple. \\
In quaternion form the group is $(\D_3\rl\Z_2;\mO\rl\V)$.
We do not pursue further in this example, which is one of a list of pseudo-simple cycles in $\R^4$ yet to be established.

\section{Discussion}\label{sec5}

We have found a complete list of finite subgroup of O(4) admitting
simple heteroclinic cycles, thus complementing the classification
initiated by \cite{km04} (cycles of types B and C) and \cite{sot03, sot05}
(homoclinic cycles). This led us to define pseudo-simple heteroclinic
cycles, a case which had not been envisaged before. An example of
a pseudo-simple cycle is given, however their classification is yet to be completed.

This work was based on the quaternionic presentation of finite subgroups of SO(4).
Note that, such an approach can be applied to other questions in equivariant bifurcation
theory in $\R^4$. Annex \ref{subgroupsO4} provides an example where a problem treated
in \cite{laumat} gets a shorter solution.
The reconstruction of the matrix group
actions, invariant planes and axes and equivariant systems with heteroclinic cycles, can be
performed from the formulas in Section \ref{sec:quaternions} and from tables
in the annexes \ref{conjugacyclasses}-\ref{Sigmaj-Deltaj}.

The subgroups of O(4) which do not admit simple heteroclinic cycles can
admit pseudo-simple heteroclinic cycle, as it is shown in subsection \ref{subsec:simple}.
A pseudo-simple cycle has at least one
equilibria $\xi_j$ where the expanding eigenvector belongs to the two-dimensional isotypic
component in the decomposition of $\Delta_j$. This implies that $L_j$ is the intersection
of several symmetric copies of $P_j$, which gives rise to a new kind of
potentially complex nearby dynamics.
Subgroups, admitting pseudo-simple heteroclinic cycles,
can be found and the cycles can be identified using the same
technique as in the present paper. In fact, the subgroups of O(4) typically
admit not just heteroclinic cycles, but more complex heteroclinic networks.
(This should be clear from the tables in annexes \ref{conjugacyclasses}-\ref{Sigmaj-Deltaj}
). Identification of such networks can be also achieved by the same approach.

According to \cite{km95a,km04,pa11}, any simple heteroclinic
cycle can be asymptotically stable, provided that eigenvalues of $df({\xi_j})$
satisfy some inequalities stated {\it ibid}. If a cycle is not asymptotically stable, it
can be stable in a weaker sense and attract a positive measure set of initial
conditions, as discussed in \cite{melb,km04,op12}.
The local extension of the basin of attraction can be described in terms of
stability indices, which were introduced in \cite{pa11}.
However this issue is beyond the scope of the present paper.

\subsection*{Acknowledgements}

The research of OP was financed in part by the grant 11-05-00167-a from
the Russian foundation for basic research, several visits to the
Observatoire de la C\^ote d'Azur (France) were
supported by the French Ministry of Higher Education and Research.

\newpage
\appendix
\section{Subgroups of O(4) that do not have one-dimensional fixed-point subspaces
\label{subgroupsO4}}

Here we give a list of subgroups of O(4) which act irreducibly and do not
possess axes of symmetry. The proof of the main theorem is based on a series
of lemmas given below.

\begin{lemma}\label{lemd1}
Consider $g\in$\,SO(4), $\Phi^{-1}g=((\cos\alpha,\sin\alpha{\bf v});(\cos\beta,\sin\beta{\bf w}))$.\\
Then $\dim\Fix<g>=2$ if and only if $\cos\alpha=\cos\beta$.
\end{lemma}

\begin{lemma}\label{lemd2}
Consider $g,s\in$\,SO(4), where
$\Phi^{-1}g=((\cos\alpha,\sin\alpha{\bf v});(\cos\alpha,\sin\alpha{\bf w}))$ and\\
$\Phi^{-1}s=((0,{\bf v});(0,{\bf w}))$.\\
Then $\Fix<g>=\Fix<s>$.
\end{lemma}

\begin{lemma}\label{lemd3}
The action of $\Gamma=\Phi(\Z_n\rl\Z_n;\Z_k\rl\Z_k)$ on $\R^4$ is
reducible.
\end{lemma}

The proofs follows from the properties of quaternions and we do not present them.

\begin{lemma}\label{lemd4}
 If a group $\Gamma\subset$\,SO(4) has one-dimensional fixed-point subspace then \\
$\Phi^{-1}\Gamma=({\bf L}\rl{\bf L}_K;{\bf R}\rl{\bf R}_K)$ satisfies
\begin{equation}\label{condl4}
{\bf L}\supset\D_s\hbox{ and }{\bf R}\supset\D_s\hbox{ for some }s\ge 2.
\end{equation}
\end{lemma}

\proof
Any one-dimensional fixed-point subspace $L$ of $\Gamma\subset$\,SO(4) is
an intersection of two isotropy planes, $P_1$ and $P_2$. Denote by $s_j$ elements of SO(4) such that
$P_j=\Fix<s_j>$. The group $<s_1,s_2>$ acting on $\R^3=\R^4\ominus L$ does not have
fixed-point subspaces, therefore $<s_1,s_2>\ne\Z_k$ for any $k$. Hence,
$<s_1,s_2>\supset\D_s$ for some $s\ge 2$,
which implies (\ref{condl4}).
\qed

\begin{lemma}\label{lemd5}
Suppose that a group $\Gamma\subset$\,SO(4) satisfies\\
(i) $\Gamma$ is not a subgroup of $(\Z_n\rl\Z_n;\Z_k\rl\Z_k)$ for any $n$ and $k$;\\
(ii) $\Gamma$ does not have one-dimensional fixed-point subspaces.\\
Then the group $\Gamma$ acts on $\R^4$ irreducibly.
\end{lemma}

\proof
There exists a group $(\Z_{rN}\rl\Z_N;\Z_{rM}\rl\Z_M)_s\subset\Gamma$
where at least one of $rN\ge3$ or $rM\ge3$ is satisfied. The elements of
$(\Z_{rN}\rl\Z_N;\Z_{rM}\rl\Z_M)_s$ act as
rotations in two absolutely perpendicular planes, $V_1$ and $V_2$. The condition (i) implies existence
of $g\in\Gamma$, such that $g\notin(\Z_{rN}\rl\Z_N;\Z_{rM}\rl\Z_M)_s$. If the action of
$\Gamma$ is reducible, then both $V_1$ and $V_2$ are $g$-invariant and $g$ acts on
both $V_1$ and $V_2$ as a reflection.
The group, generated by any $q\in(\Z_{rN}\rl\Z_N;\Z_{rM}\rl\Z_M)_s$, $q\ne e$, and
$g$, contains $(\D_s\rl\Z_1;\D_s\rl\Z_1)$ with some $s\ge2$. According to lemma \ref{lemd4},
such a group has an axis of symmetry, which contradicts (ii).
Therefore, the group $\Gamma$ acts on $\R^4$ irreducibly.
\qed

\pagebreak
\begin{theorem}\label{th3}
The following subgroups of SO(4) act on $\R^4$ irreducibly and does not have
one-dimensional fixed-point subspaces:

\hskip -2cm\begin{table}[h]
\begin{equation}\label{listth3}
\renewcommand{\arraystretch}{1.5}
\begin{array}{|l|l|l|}
\hline
(\Z_{2K_1}\rl\Z_{2K_1};\D_{K_2}\rl\D_{K_2}) &
(\D_{K_1}\rl\D_{K_1};\D_{K_2}\rl\D_{K_2}),\ K_1,K_2\hbox{ co-prime}\\
\hline
(\Z_{4K_1}\rl\Z_{2K_1};\D_{K_2}\rl\Z_{2K_2}) &
(\D_{K_1}\rl\Z_{2K_1};\D_{K_2}\rl\Z_{2K_2}),\ K_1,K_2\hbox{ co-prime}\\
\hline
(\Z_{4K_1}\rl\Z_{2K_1};\D_{2K_2}\rl\D_{K_2}) &
(\D_{2K_1}\rl\D_{K_1};\D_{K_2}\rl\Z_{2K_2}),\ K_1\hbox{ odd,} K_1,K_2\hbox{ co-prime}\\
\hline
(\Z_{2K_1}\rl\Z_{2K_1};\T\rl\T) &
(\D_{K_1}\rl\D_{K_1};\T\rl\T),\ K_1\ne 2k\\
\hline
(\Z_{6K_1}\rl\Z_{2K_1};\T\rl\V) &
(\D_{K_1}\rl\D_{K_1};\mO\rl\mO),\ K_1\ne 2k,3k\\
\hline
(\Z_{2K_1}\rl\Z_{2K_1};\mO\rl\mO) &
(\D_{K_1}\rl\Z_{2K_1};\mO\rl\T),\ K_1\ne 2k,3k\\
\hline
(\Z_{4K_1}\rl\Z_{2K_1};\mO\rl\T) &
(\D_{K_1}\rl\D_{K_1};\I\rl\I),\ K_1\ne 2k,5k\\
\hline
(\Z_{2K_1}\rl\Z_{2K_1};\I\rl\I) &
(\D_{K_1}\rl\Z_{K_1};\D_{K_2}\rl\Z_{K_2}),\ K_1,K_2\hbox{ odd, co-prime}\\
\hline
\end{array}
\end{equation}
\end{table}
\end{theorem}

The proof follows from the list of finite subgroups of SO(4) (see table \ref{listSO4}),
lemmas \ref{lemd4} and \ref{lemd5} and is not presented.

\begin{remark}
Note that the groups
$$(\Z_{2K_1}\rl\Z_{2K_1};\D_{K_2}\rl\D_{K_2})\hbox{ with }K_1\hbox{ odd, } K_1,K_2\hbox{ co-prime};\
(\Z_{2K_1}\rl\Z_{2K_1};\T\rl\T)\hbox{ with }K_1\ne2k,3k;$$
$$(\Z_{2K_1}\rl\Z_{2K_1};\mO\rl\mO)\hbox{ with }K_1\ne2k,3k\hbox{ and }
(\Z_{2K_1}\rl\Z_{2K_1};\I\rl\I)\hbox{ with }K_1\ne2k,3k,5k$$
do not have non-trivial fixed-point subspaces at all.
\end{remark}

\bigskip
\begin{lemma}\label{lemd6}
Suppose that
a finite group $\Gamma^*\subset$\,O(4), $\Gamma^*\not\subset$SO(4), acts
irreducibly in $\R^4$. Then $\Gamma^*$ possesses at least one axis of symmetry.
\end{lemma}

\proof
Recall that $\Gamma^*$ can be decomposed as
$$\Gamma^*=\Gamma\oplus\sigma\Gamma,\hbox{ where $\Gamma\subset$\,SO(4)
and $\sigma\notin$\,SO(4)}.$$
In the quaternion form
$\Phi^{-1}\Gamma=({\bf G}\rl{\bf G}_K;{\bf G}\rl{\bf G}_K)$.
If ${\bf G}\ne\Z_n$, then the existence of a one-dimensional fixed-point
subspace follows from lemma \ref{lemd4}.

Suppose that ${\bf G}=\Z_n$. Recall that $\sigma$, a reflection in $\R^4$, has $\pm1$ for
two of its eigenvalues, the other two being of the form ${\rm e}^{{\rm i}\omega}$.
If $\omega\ne k\pi$ then the reflection $\sigma$ has one-dimensional fixed-point
subspace. If $\omega=k\pi$, then it has a three-dimensional fixed-point subspace, $Q$.
As in the proof of lemma \ref{lemd5}, denote by $V_1$ and $V_2$ two invariant
subspaces of the group $\Gamma$. Since $\sigma\gamma\sigma\in\Gamma$ for
any of $\gamma\in\Gamma$, one of these subspaces belongs to $Q$, therefore
then the action of $\Gamma^*$ on $\R^4$ is reducible.
\qed

\begin{remark}
 Lauterbach and Matthews \cite{laumat} found three subgroups of SO(4) which
act irreducibly and do not have one-dimensional fixed-point subspaces.
The subgroups are denoted $G_j(m)$, where $j=1,2,3$ and $m\ge3$ is an odd integer.
In our notation, $G_1(m)$ is $(\D_4\rl\D_2;\D_m\rl\Z_{2m})$ and
$G_3(m)$ is $(\D_m\rl\D_m;\D_2\rl\D_2)$.
\end{remark}

\newpage
\section{Plane reflections for the groups listed in Table \ref{listSO4}\label{planereflections}}
We write ${\bf u}=(0,0,0,1)$, ${\bf v}_{\pm}=(0,1,\pm1,0)/\sqrt{2}$ and the permutation $\rho:~(a,b,c,d)\mapsto~(a,c,d,b)$.
\begin{table}[htdp]
{\footnotesize
\begin{center}
\begin{tabular}{|c|l|}
\hline
group ${\Gamma}$ & \mbox{Plane reflections} \\
\hline
$(\D_{2K_1}/\D_{2K_1};\D_{2K_2}/\D_{2K_2})$ & $\kappa_1(\pm)=((0,0,0,1);(0,0,0,\pm1))$ \\ $\theta_1=\pi/(2K_1)$ & $\kappa_2(n_1)=((0,\cos(n_1\theta_1),\sin(n_1\theta_1),0);(0,0,0,1))$ \\ $\theta_2=\pi/(2K_2)$ & $\kappa_3(n_2)=((0,0,0,1);(0,\cos(n_2\theta_2),\sin(n_2\theta_2),0))$ \\ & $\kappa_4(n_1,n_2)=((0,\cos(n_1\theta_1),\sin(n_1\theta_1),0);
(0,\cos(n_2\theta_2),\sin(n_2\theta_2),0))$ \\
\hline
$(\D_{2K_1r}/\Z_{4K_1};\D_{2K_2r}/\Z_{4K_2})_s$ & $\kappa_1(\pm)=((0,0,0,1);(0,0,0,\pm1))$ \\
$\theta_1^*=\theta_1/r,\ \theta_2^*=\theta_2/r$ & $\kappa_2(n_1,n_2,n_3)=((0,\cos(n_1\theta_1+n_3\theta_1^*),
\sin(n_1\theta_1+n_3\theta_1^*),0)$; \\
& ~~~~~~~~~~~~~~~~~~~~~~~~~$(0,\cos(n_2\theta_2+n_3s\theta_2^*),\cos(n_2\theta_2+n_3s\theta_2^*),0))$ \\
\hline
$(\D_{2K_1r}/\Z_{2K_1};\D_{2K_2r}/\Z_{2K_2})_s$ & $\kappa_1(n_1,n_2,n_3)=((0,\cos(2n_1\theta_1+n_3\theta_1^*),
\sin(2n_1\theta_1+n_3\theta_1^*),0);$ \\
$K_1+K_2\hbox{ odd}$ & ~~~~~~~~~~~~~~~~~~~~~$(0,\cos(2n_2\theta_2+n_3s\theta_2^*),\sin(2n_2\theta_2+n_3s\theta_2^*),0)),$ \\
& $\kappa_2(n_1,n_2,n_3)=((0,\cos((2n_1+1)\theta_1+n_3\theta_1^*), \sin((2n_1+1)\theta_1+n_3\theta_1^*),0);$ \\
& ~~~~~~~~~~~~~~~~~~$(0,\cos((2n_2+1)\theta_2+n_3s\theta_2^*), \sin((2n_2+1)\theta_2+n_3s\theta_2^*),0))$ \\
\hline
$(\D_{2K_1r}/\Z_{2K_1};\D_{2K_2r}/\Z_{2K_2})_s$ & $\kappa_1(\pm)=((0,0,0,1);(0,0,0,\pm1)),$ \\
$K_1+K_2\hbox{ even}$ & $\kappa_2(n_1,n_2,n_3)=((0,\cos(2n_1\theta_1+n_3\theta_1^*),
\sin(2n_1\theta_1+n_3\theta_1^*),0);$ \\
& ~~~~~~~~~~~~~~~~~~~~~$(0,\cos(2n_2\theta_2+n_3s\theta_2^*),\sin(2n_2\theta_2+n_3s\theta_2^*),0)),$ \\
& $\kappa_3(n_1,n_2,n_3)=((0,\cos((2n_1+1)\theta_1+n_3\theta_1^*), \sin((2n_1+1)\theta_1+n_3\theta_1^*),0);$ \\
& ~~~~~~~~~~~~~~~~~~~~~$(0,\cos((2n_2+1)\theta_2+n_3s\theta_2^*), \cos((2n_2+1)\theta_2+n_3s\theta_2^*),0))$ \\
\hline
$(\D_{2K_1}/\D_{K_1};\D_{2K_2}/\D_{K_2})$ & $\kappa_1(\pm)=((0,0,0,1);(0,0,0,\pm1))$ \\
$K_1,K_2\hbox{ even}$ & $\kappa_2(n_1)=((0,\cos(2n_1\theta_1),\sin(2n_1\theta_1),0);(0,0,0,1))$ \\
& $\kappa_3(n_2)=((0,0,0,1);(0,\cos(2n_2\theta_2),\sin(2n_2\theta_2,0)))$ \\
& $\kappa_4(n_1,n_2)=((0,\cos(2n_1\theta_1),\sin(2n_1)\theta_1),0);
(0,\cos(2n_2\theta_2),\sin(2n_2\theta_2),0))$ \\
& $\kappa_5(n_1,n_2)=((0,\cos((2n_1+1)\theta_1),\sin((2n_1+1)\theta_1),0);$ \\
& ~~~~~~~~~~~~~~~~~~~~~$(0,\cos((2n_2+1)\theta_2),\sin((2n_2+1)\theta_2),0))$ \\
\hline
$(\D_{2K_1}/\D_{K_1};\D_{2K_2}/\D_{K_2})$ & $\kappa_1(\pm)=((0,0,0,1);(0,0,0,\pm1))$ \\
$K_1,K_2\hbox{ odd}$ & $\kappa_2(n_1)=((0,\cos((2n_1+1)\theta_1),\sin((2n_1+1)\theta_1),0);(0,0,0,1))$ \\
& $\kappa_3(n_2)=((0,0,0,1);(0,\cos((2n_2+1)\theta_2),\sin((2n_2+1)\theta_2),0))$ \\
& $\kappa_4(n_1,n_2)=((0,\cos(2n_1\theta_1),\sin(2n_1\theta_1),0);
(0,\cos(2n_2\theta_2),\sin(2n_2\theta_2),0))$ \\
& $\kappa_5(n_1,n_2)=((0,\cos((2n_1+1)\theta_1),\sin((2n_1+1)\theta_1),0);$ \\
& ~~~~~~~~~~~~~~~~~~~~~$(0,\cos((2n_2+1)\theta_2),\sin((2n_2+1)\theta_2),0))$ \\
\hline
$(\D_{2K_1}/\D_{K_1};\D_{2K_2}/\D_{K_2})$ & $\kappa_1(n_1)=((0,\cos((2n_1+1)\theta_1),\sin((2n_1+1)\theta_1),0);(0,0,0,1))$ \\
$K_1\hbox{ even}$, $K_2\hbox{ odd}$ & $\kappa_2(n_2)=((0,0,0,1);(0,\cos(2n_2\theta_2),\sin(2n_2\theta_2),0))$ \\
 & $\kappa_3(n_1,n_2)=((0,\cos(2n_1\theta_1),\sin(2n_1\theta_1),0);
(0,\cos(2n_2\theta_2),\sin(2n_2\theta_2),0))$ \\
& $\kappa_4(n_1,n_2)=((0,\cos((2n_1+1)\theta_1),\sin((2n_1+1)\theta_1),0);$ \\
&  ~~~~~~~~~~~~~~~~~~~~~$(0,\cos((2n_2+1)\theta_2),\sin((2n_2+1)\theta_2),0))$\\
\hline
$(\D_{2K_1}/\D_{K_1};\D_{2K_2}/\Z_{4K_2})$ & $\kappa_1(\pm)=((0,0,0,1);(0,0,0,\pm1))$ \\
$K_1\hbox{ even}$ & $\kappa_2(n_1)=((0,\cos(2n_1\theta_1),\sin(2n_1\theta_1),0);(0,0,0,1))$ \\
& $\kappa_3(n_1,n_2)=((0,\cos((2n_1+1)\theta_1),\sin((2n_1+1)\theta_1),0);$ \\
& ~~~~~~~~~~~~~~~~~~~~~$(0,\cos(n_2\theta_2),\sin(n_2\theta_2),0))$ \\
\hline
\end{tabular}
\end{center}
}
\end{table}
\newpage

\noindent {\bf Annex \ref{planereflections} continued.} \\
We write ${\bf w}_{\pm\pm}=(0,1,\pm\tau\pm\tau^{-1})/2$ and ${\bf w}_{\pm\pm}^*=(0,1,\pm\tau^*,\pm(\tau^*)^{-1})/2$, where $\tau=(\sqrt{5}+1)/2$ and $\tau^*=(-\sqrt{5}+1)/2$.
\begin{table}[htdp]
{\footnotesize
\begin{center}
\begin{tabular}{|c|l|}
\hline
group ${\Gamma}$ & \mbox{Plane reflections} \\
\hline
$(\D_{2K_1}/\D_{K_1};\D_{2K_2}/\Z_{4K_2})$ & $\kappa_1(n_1)=((0,\cos(2n_1\theta_1),\sin(2n_1\theta_1),0);(0,0,0,1))$ \\
$K_1\hbox{ odd}$ & $\kappa_2(n_2)=((0,0,0,1);(0,\cos(n_2\theta_2),\sin(n_2\theta_2),0))$ \\
& $\kappa_3(n_1,n_2)=((0,\cos((2n_1+1)\theta_1),\sin((2n_1+1)\theta_1),0);$ \\
& ~~~~~~~~~~~~~~~~~~~~~$(0,\cos(n_2\theta_2),\sin(n_2\theta_2),0))$ \\
\hline
$(\D_{2K}/\D_{2K};\T/\T)$ & $\kappa_1(\pm,r)=((0,0,0,\pm1);\rho^r{\bf u}),\ r=0,1,2$ \\
$\theta=\pi/(2K)$ & $\kappa_2(n,r)=((0,\cos(n\theta),\sin(n\theta),0);\rho^r{\bf u})$ \\
\hline
$(\D_{2K}/\D_{2K};\mO/\mO)$ & $\kappa_1(\pm,r)=((0,0,0,\pm1);\rho^r{\bf u})$,
$\kappa_2(\pm,r,\pm)=((0,0,0,\pm1);\rho^r{\bf v}_{\pm})$ \\
& $\kappa_3(n,r)=((0,\cos(n\theta),\sin(n\theta),0);\rho^r{\bf u})$ \\
& $\kappa_4(n,r,\pm)=((0,\cos(n\theta),\sin(n\theta),0);\rho^r{\bf v}_{\pm})$ \\
\hline
$(\D_{2K}/\Z_{4K};\mO/\T)$ & $\kappa_1(\pm,r)=((0,0,0,\pm1);\rho^r{\bf u})$ \\
& $\kappa_2(n,r,\pm)=((0,\cos(n\theta),\sin(n\theta),0);\rho^r{\bf v}_{\pm})$ \\
\hline
$(\D_{2K}/\D_K;\mO/\T)$& $\kappa_1(\pm,r)=((0,0,0,\pm1);\rho^r{\bf u})$ \\
$K\hbox{ even}$& $\kappa_2(n,r)=((0,\cos(2n\theta),\sin(2n\theta),0);\rho^r{\bf u})$ \\
& $\kappa_3(n,r,\pm)=((0,\cos((2n+1)\theta),\sin((2n+1)\theta),0);\rho^r{\bf v}_{\pm})$ \\
\hline
$(\D_{2K}/\D_K;\mO/\T)$ & $\kappa_1(\pm,r,\pm)=((0,0,0,\pm1);\rho^r{\bf v}_{\pm})$ \\
$K\hbox{ odd}$ & $\kappa_2(n,r)=((0,\cos(2n\theta),\sin(2n\theta),0);\rho^r{\bf u})$ \\
& $\kappa_3(n,r,\pm)=((0,\cos((2n+1)\theta),\sin((2n+1)\theta),0);\rho^r{\bf v}_{\pm})$ \\
\hline
$(\D_{6K}/\Z_{4K};\mO/\V)$ & $\kappa_1(\pm,r)=((0,0,0,\pm1);\rho^r{\bf u}))$ \\
$\theta=\pi/(6K)$ & $\kappa_2(n,\pm)=((0,\cos(3n\theta),\sin(3n\theta),0);{\bf v}_{\pm})$ \\
& $\kappa_3(n,\pm)=((0,\cos(3n+1)\theta),\sin((3n+1)\theta),0);\rho{\bf v}_{\pm})$ \\
& $\kappa_4(n,\pm)=((0,\cos(3n+2)\theta),\sin((3n+2)\theta),0);\rho^2{\bf v}_{\pm})$ \\
\hline
$(\D_{2K}/\D_{2K};\I/\I)$ & $\kappa_1(\pm,r)=((0,0,0,\pm1);\rho^r{\bf u}),\
\kappa_1'(\pm,r,\pm\pm)=((0,0,0,\pm1);\rho^r{\bf w}_{\pm\pm})$ \\
$\theta=\pi/(2K)$ & $\kappa_2(n,r)=((0,\cos(n\theta),\sin(n\theta),0);\rho^r{\bf u})$ \\
& $\kappa_2'(n,r,\pm\pm)=((0,\cos(n\theta),\sin(n\theta),0);\rho^r{\bf w}_{\pm\pm})$ \\
\hline
$(\T/\T;\T/\T)$ & $\kappa_1(\pm,r,s)=\pm(\rho^r{\bf u};\rho^s{\bf u})$ \\
\hline
$(\T/\Z_2;\T/\Z_2)$ & $\kappa_1(\pm,r)=\pm(\rho^r{\bf u};\rho^r{\bf u})$ \\
\hline
$(\T/\V;\T/\V)$& $\kappa_1(\pm,r,s)=\pm(\rho^r{\bf u};\rho^s{\bf u})$ \\
\hline
$(\T/\T;\mO/\mO)$ & $\kappa_1(\pm,r,s)=\pm(\rho^r{\bf u};\rho^s{\bf u}),\
\kappa_2(\pm,r,s,\pm)=\pm(\rho^r{\bf u};\rho^s{\bf v}_{\pm})$ \\
\hline
$(\T/\T;\I/\I)$ & $\kappa_1(\pm,r,s)=\pm(\rho^r{\bf u};\rho^s{\bf u}),\
\kappa_1'(\pm,r,s,\pm\pm)=\pm(\rho^r{\bf u};\rho^s{\bf w}_{\pm\pm})$ \\
\hline
$(\mO/\mO;\mO/\mO)$ & $\kappa_1(\pm,r,s)=\pm(\rho^r{\bf u};\rho^s{\bf u}),\
\kappa_2(\pm,r,s,\pm)=\pm(\rho^r{\bf u};\rho^s{\bf v}_{\pm})$ \\
& $\kappa_3(\pm,r,\pm,s)=\pm(\rho^r{\bf v}_{\pm};\rho^s{\bf u}),\
\kappa_4(\pm,r,\pm,s,\pm)=\pm(\rho^r{\bf v}_{\pm};\rho^s{\bf v}_{\pm})$ \\
\hline
$(\mO/\Z_2;\mO/\Z_2)$ & $\kappa_1(\pm,r)=\pm(\rho^r{\bf u};\rho^r{\bf u}),\
\kappa_2(\pm,r,\pm)=\pm(\rho^r{\bf v}_{\pm};\rho^r{\bf v}_{\pm})$ \\
\hline
$(\mO/\V;\mO/\V)$ & $\kappa_1(\pm,r,s)=\pm(\rho^r{\bf u};\rho^s{\bf u}),\
\kappa_2(\pm,r,\pm,\pm)=\pm(\rho^r{\bf v}_{\pm};\rho^r{\bf v}_{\pm})$ \\
\hline
$(\mO/\T;\mO/\T)$& $\kappa_1(\pm,r,s)=\pm(\rho^r{\bf u};\rho^s{\bf u}),\
\kappa_2(\pm,r,\pm,s,\pm)=\pm(\rho^r{\bf v}_{\pm};\rho^s{\bf v}_{\pm})$ \\
\hline
$(\mO/\mO;\I/\I)$& $\kappa_1(\pm,r,s)=\pm(\rho^r{\bf u};\rho^s{\bf u}),\
\kappa_1'(\pm,r,s,\pm\pm)=\pm(\rho^r{\bf u};\rho^s{\bf w}_{\pm\pm})$ \\
& $\kappa_2(\pm,r,\pm,s)=\pm(\rho^r{\bf v}_{\pm};\rho^s{\bf u}),\
\kappa_2'(\pm,r,\pm,s,\pm\pm)=(\rho^r{\bf v}_{\pm};\rho^s{\bf w}_{\pm\pm})$ \\
\hline
\end{tabular}
\end{center}
}
\end{table}

\newpage
\noindent {\bf Annex \ref{planereflections} continued.} \\
We write ${\bf w}_{\pm\pm}=(0,1,\pm\tau\pm\tau^{-1})/2$ and ${\bf w}_{\pm\pm}^*=(0,1,\pm\tau^*,\pm(\tau^*)^{-1})/2$, where $\tau=(\sqrt{5}+1)/2$ and $\tau^*=(-\sqrt{5}+1)/2$.
\begin{table}[htdp]
{\footnotesize
\begin{center}
\begin{tabular}{|c|l|}
\hline
group ${\Gamma}$ & \mbox{Plane reflections} \\
\hline
$(\I/\I;\I/\I)$ & $\kappa_1(\pm,r,s)=\pm(\rho^r{\bf u};\rho^s{\bf u}),\
\kappa_1'(\pm,r,s,\pm\pm)=\pm(\rho^r{\bf u};\rho^s{\bf w}_{\pm\pm})$ \\
& $\kappa_1''(\pm,r,\pm\pm,s)=\pm(\rho^r{\bf w}_{\pm\pm};\rho^s{\bf u})$ \\
& $\kappa_1'''(\pm,r,\pm\pm,s,\pm\pm)=\pm(\rho^r{\bf w}_{\pm\pm};\rho^s{\bf w}_{\pm\pm})$ \\
\hline
$(\I/\Z_2;\I/\Z_2)$& $\kappa_1(\pm,r)=\pm(\rho^r{\bf u};\rho^r{\bf u}),\
\kappa_1'(\pm,r,\pm\pm)=\pm(\rho^r{\bf w}_{\pm\pm};\rho^r{\bf w}_{\pm\pm})$ \\
\hline
$(\I^{\dagger}/\Z_2;\I/\Z_2)$ & $\kappa_1(\pm,r)=\pm(\rho^r{\bf u};\rho^r{\bf u}),\
\kappa_1'(\pm,r,\pm\pm)=\pm(\rho^r{\bf w}_{\pm\pm}^*;\rho^r{\bf w}_{\pm\pm})$ \\
\hline
$(\D_{2rK_1}/\Z_{K_1};\D_{2rK_2}/\Z_{K_2})_s$ & $\kappa_1=((0,0,0,1);(0,0,0,1))$ \\
$K_1,K_2\hbox{ odd}$ & $\kappa_2(n_1,n_2,n_3)=((0,\cos(2n_1\theta_1+n_3\theta_1^*),\sin(2n_1\theta_1+n_3\theta_1^*),0)$ \\
$\theta_1=\pi/K_1,\theta_2=\pi/K_2$ & $(0,\cos(2n_2\theta_2+sn_3\theta_2^*),\cos(2n_2\theta_2+sn_3\theta_2^*),0))$ \\
$\theta_1^*=\theta_1/(2r),\theta_2^*=\theta_2/(2r)$ & $\kappa_3(n_1,n_2,n_3)=((0,\cos((2n_1+1)\theta_1+n_3\theta_1^*),\sin((2n_1+1)\theta_1+n_3\theta_1^*),0);$ \\
& $(0,\cos((2n_2+1)\theta_2+sn_3\theta_2^*),\sin((2n_2+1)\theta_2+sn_3\theta_2^*),0))$ \\
\hline
$(\T/\Z_1;\T/\Z_1)$ & $\kappa(r)=(\rho^r{\bf u};\rho^r{\bf u})$ \\
\hline
$(\mO/\Z_1;\mO/\Z_1)$ & $\kappa_1(r)=(\rho^r{\bf u};\rho^r{\bf u}),\
\kappa_2(r,\pm)=(\rho^r{\bf v}_{\pm};\rho^r{\bf v}_{\pm})$ \\
\hline
$(\mO/\Z_1;\mO/\Z_1)^{\dagger}$ & $\kappa_1(r)=(\rho^r{\bf u};\rho^r{\bf u}),\
\kappa_2(r,\pm)=-(\rho^r{\bf v}_{\pm};\rho^r{\bf v}_{\pm})$ \\
\hline
$(\I/\Z_1;\I/\Z_1)$ & $\kappa_1(r)=(\rho^r{\bf u};\rho^r{\bf u}),\
\kappa_1'(r,\pm\pm)=(\rho^r{\bf w}_{\pm\pm};\rho^r{\bf w}_{\pm\pm})$ \\
\hline
$(\I/\Z_1^{\dagger};\I/\Z_1)^{\dagger}$ & $\kappa_1(r)=(\rho^r{\bf u};\rho^r{\bf u}),\
\kappa_1'(r,\pm\pm)=(\rho^r{\bf w}_{\pm\pm}^*;\rho^r{\bf w}_{\pm\pm})$ \\
\hline
\end{tabular}
\end{center}
}
\end{table}

\section{Conjugacy classes of isotropy subgroups of finite groups $\Gamma$ satisfying
$\dim\Fix\,(\Sigma)=2$ and $\dim\Fix\,(\Delta)=1$\label{conjugacyclasses}}

We list all such groups $\Sigma$, they always satisfy
$\Sigma\cong\Z_2$. Only selected $\Delta$ are given: we list all
$\Delta\cong(\Z_2)^2$; for some of $\Delta\not\cong(\Z_2)^2$ we indicate
plane reflections, which are elements of the groups.
{\small
\begin{sidewaystable}
$$
\begin{array}{|l|l|l|}
\hline
{\Gamma} & \Sigma & \Delta \\
\hline
(\D_{2K_1}/\D_{2K_1};\D_{2K_2}/\D_{2K_2})&\{e,\kappa_1(\pm)\};&
\{e,\kappa_2(n_1),\kappa_3(n_2),\kappa_4(n_1-K_1,n_2+K_2)\}:\\
&\{e,\kappa_2(n_1)\}:\ n_1\hbox{ even or odd;}&
n_1\hbox{ even or odd, }n_2\hbox{ even or odd}\\
&\{e,\kappa_3(n_2)\}:\ n_2\hbox{ even or odd;}&\\
&\{e,\kappa_4(n_1,n_2)\}:\ n_1\hbox{ even or odd,}&\\
&n_2\hbox{ even or odd}&\\
\hline
(\D_{2K_1r}/\Z_{4K_1};\D_{2K_2r}/\Z_{4K_2})_s&
\{e,\kappa_1(+)\};\ \{e,\kappa_1(-)\};&
\{e,\kappa_1(+),\kappa_2(n_1,n_2,n_3),\kappa_2(n_1+K_1,n_2+K_2,n_3)\}:\\
K_1,K_2\hbox{ odd},&\{e,\kappa_2(n_1,n_2,n_3)\}:&n_1+n_2\hbox{ even or odd;}\\
K_1,K_2\hbox{ co-prime},&n_1+n_3\hbox{ even or odd,}&
\{e,\kappa_1(-),\kappa_2(n_1,n_2,n_3),\kappa_2(n_1+K_1,n_2+K_2,n_3)\}:\\
r\hbox{ odd}&n_2+n_3\hbox{ even or odd}&n_1+n_2\hbox{ even or odd}\\
\hline
(\D_{2K_1r}/\Z_{4K_1};\D_{2K_2r}/\Z_{4K_2})_s&
\{e,\kappa_1(+)\};\ \{e,\kappa_1(-)\};&
\{e,\kappa_1(+),\kappa_2(n_1,n_2,n_3),\kappa_2(n_1+K_1,n_2+K_2,n_3)\}:\
n_3\hbox{ even or odd;}\\
K_1,K_2\hbox{ odd},&\{e,\kappa_2(n_1,n_2,n_3)\}:&
\{e,\kappa_1(-),\kappa_2(n_1,n_2,n_3),\kappa_2(n_1+K_1,n_2+K_2,n_3)\}:\
n_3\hbox{ even or odd}\\
K_1,K_2\hbox{ co-prime},&n_1+n_2\hbox{ even or odd, }&\\
r\hbox{ even}&n_3\hbox{ even or odd}&\\
\hline
(\D_{2K_1r}/\Z_{4K_1};\D_{2K_2r}/\Z_{4K_2})_s&
\{e,\kappa_1(+)\};\ \{e,\kappa_1(-)\};&
\{e,\kappa_1(+),\kappa_2(n_1,n_2,n_3),\kappa_2(n_1+K_1,n_2+K_2,n_3)\}:\\
K_1\hbox{ even}, K_1,K_2\hbox{ co-prime},&
\{e,\kappa_2(n_1,n_2,n_3)\}:&n_1+n_3\hbox{ even or odd;}\\
r\hbox{ odd}&n_1+n_3\hbox{ even or odd,}&
\{e,\kappa_1(-),\kappa_2(n_1,n_2,n_3),\kappa_2(n_1+K_1,n_2+K_2,n_3)\}:\\
&n_2+n_3\hbox{ even or odd}&
n_1+n_3\hbox{ even or odd}\\
\hline
(\D_{2K_1r}/\Z_{2K_1};\D_{2K_2r}/\Z_{2K_2})_s&
\{e,\kappa_1(+)\};\ \{e,\kappa_1(-)\};&\{e,\kappa_1(+),\kappa_2(n_1,n_2,n_3),
\kappa_3(n_1+(K_1-1)/2,n_2+(K_2-1)/2,n_3)\}:\\
K_1,K_2\hbox{ odd},&\{e,\kappa_2(n_1,n_2,n_3)\}:&n_3\hbox{ even or odd};\\
K_1,K_2\hbox{ co-prime}&n_3\hbox{ even or odd};&
\{e,\kappa_1(-),\kappa_2(n_1,n_2,n_3),\kappa_3(n_1+(K_1-1)/2,n_2+(K_2-1)/2,n_3)\}:\\
&\{e,\kappa_3(n_1,n_2,n_3)\}:&n_3\hbox{ even or odd}\\
&n_3\hbox{ even or odd}&\\
\hline
\end{array}
$$
\end{sidewaystable}

\begin{sidewaystable}
$$
\begin{array}{|l|l|l|}
\hline
{\Gamma} & \Sigma & \Delta\\
\hline
(\D_{2K_1}/\D_{K_1};\D_{2K_2}/\D_{K_2})&\{e,\kappa_1(\pm)\};\
\{e,\kappa_2(n_1)\};\ \{e,\kappa_3(n_2)\};&
\{e,\kappa_2(n_1),\kappa_3(n_2),\kappa_4(n_1-K_1/2,n_2+K_2/2)\}:\\
K_1,K_2\hbox{ even}&\{e,\kappa_4(n_1,n_2)\}:&
n_1+n_2\hbox{ even or odd}\\
&n_1+n_2\hbox{ even or odd};&\\
&\{e,\kappa_5(n_1,n_2)\}:&\\
&n_1+n_2\hbox{ even or odd}&\\
\hline
(\D_{2K_1}/\D_{K_1};\D_{2K_2}/\D_{K_2})&\{e,\kappa_1(\pm)\};\
\{e,\kappa_2(n_1)\};\ \{e,\kappa_3(n_1)\};&
\{e,\kappa_2(n_1),\kappa_3(n_2),\kappa_4(n_1-(K_1-1)/2,n_2+(K_2+1)/2)\}:\\
K_1,K_2\hbox{ odd}&\{e,\kappa_4(n_1,n_2)\}:&n_1+n_2\hbox{ even or odd}\\
&n_1+n_2\hbox{ even or odd;}&\\
&\{e,\kappa_5(n_1,n_2)\}:&\\
&n_1+n_2\hbox{ even or odd}&\\
\hline
(\D_{2K_1}/\D_{K_1};\D_{2K_2}/\D_{K_2})&\{e,\kappa_1\};\ \{e,\kappa_2(n_1)\};&
\{e,\kappa_1(n_1),\kappa_2(n_2),\kappa_4(n_1-K_1/2,n_2+(K_2+1)/2)\}:\\
K_1\hbox{ even,}&\{e,\kappa_3(n_1,n_2)\}:\ n_1+n_2\hbox{ even or odd;}&n_1+n_2\hbox{ even or odd}\\
K_2\hbox{ odd}&\{e,\kappa_4(n_1,n_2)\}:\ n_1+n_2\hbox{ even or odd}&\\
\hline
(\D_{2K_1}/\D_{K_1};\D_{2K_2}/\Z_{4K_2})&\{e,\kappa_1\};\ \{e,\kappa_2\};&
\{e,\kappa_1((-1)^s),\kappa_3(n_1,n_2),\kappa_3(n_1+K_1/2,n_2+K_2\}:\\
K_1,K_2\hbox{ co-prime},&\{e,\kappa_3(n_1,n_2)\}:\ n_2\hbox{ even or odd}&s+n_1\hbox{ even or odd}\\
K_1\hbox{ even}&&\\
\hline
(\D_{2K_1}/\D_{K_1};\D_{2K_2}/\Z_{4K_2})&
\{e,\kappa_1(n_1)\}:\ n_1\hbox{ even or odd;}&
\{e,\kappa_1(n_1),\kappa_2(n_2),\kappa_3(n_1-(K_1-1)/2,n_2+K_2)\}:\\
K_1\hbox{ odd}&\{e,\kappa_2(n_1)\}:\ n_2\hbox{ even or odd;}&
n_1\hbox{ even or odd, }n_2\hbox{ even or odd}\\
&\{e,\kappa_3(n_1,n_2)\}:\ n_2\hbox{ even or odd}&\\
\hline
(\D_{2K}/\D_{2K};\T/\T)&\{e,\kappa_1(\pm,r)\};&
\{e,\kappa_1(\pm,r),\kappa_2(n,r+1),\kappa_2(n+K,r+2)\}:\\
&\{e,\kappa_2(n,r)\}:\ n\hbox{ even or odd}&n\hbox{ even or odd}\\
\hline
(\D_{2K}/\D_{2K};\mO/\mO)&\{e,\kappa_1\};\ \{e,\kappa_2\};&
\{e,\kappa_1(\pm,r),\kappa_3(n,r+1),\kappa_3(n+K,r+2)\};\\
K\hbox{ odd}&\{e,\kappa_3(n,r)\}:\ n\hbox{ even or odd;}&
\{e,\kappa_1(\pm,r),\kappa_4(n,r,\pm),\kappa_4(n+K,r,\mp)\};\\
&\{e,\kappa_4(n,r,\pm)\}:\ n\hbox{ even or odd}&
\{e,\kappa_2(\pm,r,\pm),\kappa_3(n,r),\kappa_4(n+K,r,\mp)\}:\
n\hbox{ even or odd}\\
\hline
\end{array}
$$
Continuation of Annex \ref{conjugacyclasses}.
\end{sidewaystable}

\begin{sidewaystable}
$$
\begin{array}{|l|l|l|}
\hline
{\Gamma} & \Sigma & \Delta\\
\hline
(\D_{2K}/\Z_{4K};\mO/\T)&\{e,\kappa_1(\pm,r)\};&
\{e,\kappa_1(\pm,r),\kappa_2(n,r,+),\kappa_2(n\pm K,r,-)\}:\\
K\hbox{ even}&\{e,\kappa_2(n,r,\pm)\}:\ n\hbox{ even or odd}&n\hbox{ even or odd}\\
\hline
(\D_{2K}/\Z_{4K};\mO/\T)&\{e,\kappa_1(\pm,r)\};&
\{e,\kappa_1((-1)^s,r),\kappa_2(n,r,(-1)^s),\kappa_2(n+(-1)^sK,r,(-1)^{s+1})\}:\\
K\hbox{ odd}&\{e,\kappa_2(n,r,\pm)\}:\ n\hbox{ even or odd}&n+s\hbox{ even or odd}\\
\hline
(\D_{2K}/\D_K;\mO/\T)&\{e,\kappa_1(\pm,r)\};&
\{e,\kappa_1(\pm,r),\kappa_2(n,r+1),\kappa_2(n+K/2,r+2)\};\\
K=4\tilde k&\{e,\kappa_2(n,r)\};\ \{e,\kappa_3(n,r,\pm)\}&
\{e,\kappa_1(\pm,r),\kappa_3(n,r,\pm),\kappa_3(n-K/2,r,\mp)\}\\
\hline
(\D_{2K}/\D_K;\mO/\T)&\{e,\kappa_1(\pm,r,\pm)\};&
\{e,\kappa_1(\pm,r,(-1)^s),\kappa_2(n,r),\kappa_3(n+(K+1)/2,r,(-1)^{s+1})\}:\\
K\hbox{ odd}&\{e,\kappa_2(n,r\};\ \{e,\kappa_3(n,r,\pm\}&n+s\hbox{ even or odd}\\
\hline
(\D_{6K}/\Z_{4K};\mO/\V)&\{e,\kappa_1(\pm,r)\};&
\{e,\kappa_1(\pm,r),\kappa_{r+2}(n,\pm),
\kappa_{r+2}(n+K,\mp)\}:\ n\hbox{ even or odd}\\
K\hbox{ even}&\{e,\kappa_j(n,\pm\},\ j=2,3,4:&\\
&n\hbox{ even or odd}&\\
\hline
(\D_{6K}/\Z_{4K};\mO/\V)&\{e,\kappa_1(\pm,r)\};&
\{e,\kappa_1((-1)^s,r),\kappa_{r+2}(n,(-1)^s),
\kappa_{r+2}(n+K,(-1)^{s+1})\}:\\
K\hbox{ odd}&\{e,\kappa_j(n,\pm\},\ j=2,3,4:&n+s\hbox{ even or odd}\\
&n\hbox{ even or odd}&\\
\hline
(\D_{2K}/\D_{2K};\I/\I)&\{e,\kappa_1\},\ \{e,\kappa_1'\};&
\{e,\kappa_1(\pm,r),\kappa_2(n,r+1),\kappa_2(n+K,r+2)\}:\\
&\{e,\kappa_2(n)\},\ \{e,\kappa_2'(n)\}:&n\hbox{ even or odd}\\
&n\hbox{ even or odd}&\\
\hline
(\T/\T;\T/\T)&\{e,\kappa_1\}&
\kappa_1(\pm,r,s),\kappa_1(\pm,r+1,s+1),\kappa_1(\pm,r+2,s+2)\in\Delta_1;\\
&&\kappa_1(\pm,r,s),\kappa_1(\pm,r+2,s+1),\kappa_1(\pm,r+1,s+2)\in\Delta_2\\
\hline
(\T/\Z_2;\T/\Z_2)&\{e,\kappa_1(+,r)\};\ \{e,\kappa_1(-,r)\}&
\kappa_1(\pm,r),\kappa_1(\pm,r+1),\kappa_r(\pm,r+2)\in\Delta_1;\\
&&\{e,\kappa_1(\pm,r),\kappa_1(\mp,r+1),\kappa_r(\mp,r+2)\}\\
\hline
(\T/\V;\T/\V)&\{e,\kappa_1(r,r)\};\ \{e,\kappa_1(r,r+1)\};&
\kappa_1(\pm,r,r),\kappa_1(\pm,r+1,r+1),\kappa_1(\pm,r+2,r+2)\in\Delta_1;\\
&\{e,\kappa_1(r,r+2)\}&
\kappa_1(\pm,r,r+1),\kappa_1(\pm,r+1,r+2),\kappa_1(\pm,r+2,r)\in\Delta_2;\\
&&\kappa_1(\pm,r,r+2),\kappa_1(\pm,r+1,r),\kappa_1(\pm,r+2,r+1)\in\Delta_3;\\
&&\{e,\kappa_1(\pm,r,r),\kappa_1(\pm,r+1,r+2),\kappa_1(\pm,r+2,r+1)\}\\
\hline
(\T/\T;\mO/\mO)&\{e,\kappa_1\};\ \{e,\kappa_2\}&
\kappa_1(\pm,r,s),\kappa_1(\pm,r+1,s+1+t),\kappa_1(\pm,r+2,s+2-t)\in\Delta_1,\ t=0,1;\\
&&\{e,\kappa_1(\pm,r,s),\kappa_2(\pm,r+1+t,s,\pm),\kappa_2(\pm,r+2-t,s,\mp)\},\ t=0,1\\
\hline
(\T/\T;\I/\I)&\{e,\kappa_1\},\ \{e,\kappa_1'\}&
\kappa_1(\pm,r,s),\kappa_1(\pm,r+1,s+1),\kappa_1(\pm,r+2,s+2)\in\Delta_1;\\
&&\kappa_1(\pm,r,s),\kappa_1(\pm,r+2,s+1),\kappa_1(\pm,r+1,s+2)\in\Delta_2\\
\hline
(\D_{2rK_1}/\Z_{K_1};\D_{2rK_2}/\Z_{K_2})&
\{e,\kappa_1\};\ \{e,\kappa_2\};\ \{e,\kappa_3\}&
\{e,\kappa_1,\kappa_2(n_1,n_2),\kappa_3(n_1+(K_1-1)/2,n_2+(K_2-1)/2)\}:\\
K_1,K_2\hbox{ odd, co-prime}&&n_1+n_2\hbox{ even or odd} \\
\hline
(\T/\Z_1;\T/\Z_1)&\{e,\kappa_1(r)\}&
\kappa_1(r),\kappa_1(r+1),\kappa_1(r+2)\in\Delta_1\\
\hline
\end{array}
$$
Continuation of annex \ref{conjugacyclasses}.
\end{sidewaystable}
}

\newpage

\section{Pairs $\Sigma_j, \Delta_j$ satisfying conditions I-II-III in section
\ref{mainproofs}\label{Sigmaj-Deltaj}}

In each case, the sequence $j=1,\dots,m$ defines the building block. Note that,
in all cases $m\leq 4$. The symmetry $\gamma$ is such that
$\Sigma_m\subset\Delta_m\cap\gamma\Delta_1\gamma^{-1}$, hence insuring the
existence of a cycle of heteroclinic connections (see condition {\bf C3} in Lemma
\ref{lem12}).

\begin{sidewaystable}

$$
\begin{array}{|l|l|l|}
\hline
{\Gamma} & \Sigma_j,\Delta_j\mbox{ and }\gamma & \alpha_j\hbox{ and }\beta_j\\
\hline
(\D_{2K_1}/\D_{2K_1};\D_{2K_2}/\D_{2K_2})&
\Sigma_1=\{e,\kappa_2(0)\},\ \Sigma_2=\{e,\kappa_3(1)\}&\alpha_1=\theta_1/2,\
\beta_1=\pi-\alpha_1\\
&\Sigma_3=\{e,\kappa_2(1)\},\ \Sigma_4=\{e,\kappa_3(0)\}&\alpha_2=\theta_2/2,\
\beta_2=\pi-\alpha_2\\
&\Delta_1=\{e,\kappa_2(0),\kappa_3(0),\kappa_4(-K_1,K_2)\},&
\alpha_3=\theta_1/2,\ \beta_3=\pi-\alpha_3\\
&\Delta_2=\{e,\kappa_2(0),\kappa_3(1),\kappa_4(-K_1,K_2+1)\}&
\alpha_4=\theta_2/2,\ \beta_4=\pi-\alpha_4\\
&\Delta_3=\{e,\kappa_2(1),\kappa_3(1),\kappa_4(-K_1+1,K_2+1)\},&\\
&\Delta_4=\{e,\kappa_2(1),\kappa_3(0),\kappa_4(-K_1+1,K_2)\};\ \gamma=e&\\
\hline
(\D_{2K_1r}/\Z_{4K_1};\D_{2K_2r}/\Z_{4K_2})_s&
\Sigma_1=\{e,\kappa_2(0,0,0)\},\ \Sigma_2=\{e,\kappa_2(K_1,K_2,0)\}&
\alpha_1=\pi/2,\ \beta_1=\pi/2\\
K_1,K_2\hbox{ co-prime}&
\Delta_1=\{e,\kappa_1(-),\kappa_2(0,0,0),\kappa_2(K_1,3K_2,0)\},&
\alpha_2=\pi/2,\ \beta_2=\pi/2\\
&\Delta_2=\{e,\kappa_1(+),\kappa_2(0,0,0),\kappa_2(K_1,K_2,0)\};&\\
&\gamma=((1,0,0,0);(0,0,0,1))&\\
\hline
(\D_{2K_1r}/\Z_{2K_1};\D_{2K_2r}/\Z_{2K_2})_s&
\Sigma_1=\{e,\kappa_2(0,0,0)\},\ \Sigma_2=\{e,\kappa_3((K_1-1)/2,(K_2-1)/2,0)\}&
\alpha_1=\pi/2,\ \beta_1=\pi/2\\
K_1,K_2\hbox{ odd, co-prime}&
\Sigma_3=\{e,\kappa_2(0,K_2,0)\},\ \Sigma_4=\{e,\kappa_3((K_1-1)/2,(3K_2-1)/2,0)\}&
\alpha_2=\pi/2,\ \beta_2=\pi/2\\
&\Delta_1=\{e,\kappa_1(-),\kappa_2(0,0,0),\kappa_3((K_1-1)/2,(3K_2-1)/2,0)\}&
\alpha_3=\pi/2,\ \beta_3=\pi/2\\
&\Delta_2=\{e,\kappa_1(+),\kappa_2(0,0,0),\kappa_3((K_1-1)/2,(K_2-1)/2,0)\}&
\alpha_4=\pi/2,\ \beta_4=\pi/2\\
&\Delta_3=\{e,\kappa_1(+),\kappa_2(0,2K_2,0),\kappa_3((K_1-1)/2,(K_2-1)/2,0)\}&\\
&\Delta_4=\{e,\kappa_1(+),\kappa_2(0,2K_2,0),\kappa_3((K_1-1)/2,(3K_2-1)/2,0)\},\
\gamma=e&\\
\hline
(\D_{2K_1}/\D_{K_1};\D_{2K_2}/\D_{K_2})&
\Sigma_1=\{e,\kappa_3(0)\},\ \Sigma_2=\{e,\kappa_2(1)\},&
\alpha_1=\theta_1,\ \beta_1=\pi-\alpha_1\\
K_1,K_2\hbox{ even}&\Delta_1=\{e,\kappa_2(0),\kappa_3(0),\kappa_4(-K_1/2,K_2/2)\},&\\
&\Delta_2=\{e,\kappa_2(1),\kappa_3(0),\kappa_4(K_1/2,K_2)\},&
\alpha_2=\theta_2,\ \beta_2=\pi-\alpha_2\\
&\gamma=((\cos\theta_1,0,0,\sin\theta_1);(\cos\theta_2,0,0,\sin\theta_2)&\\
\hline
(\D_{2K_1}/\D_{K_1};\D_{2K_2}/\D_{K_2})&
\Sigma_1=\{e,\kappa_3(0)\},\ \Sigma_2=\{e,\kappa_2(1)\},&
\alpha_1=\theta_1,\ \beta_1=\pi-\alpha_1\\
K_1,K_2\hbox{ odd}&
\Delta_1=\{e,\kappa_2(0),\kappa_3(0),\kappa_4(-(K_1-1)/2,(K_2-1)/2)\},&\\
&\Delta_2=\{e,\kappa_2(1),\kappa_3(0),\kappa_4((K_1-1)/2,(K_2-1)/2)\},&
\alpha_2=\theta_2,\ \beta_2=\pi-\alpha_2\\
&\gamma=((\cos\theta_1,0,0,\sin\theta_1);(\cos\theta_2,0,0,\sin\theta_2)&\\
\hline
(\D_{2K_1}/\D_{K_1};\D_{2K_2}/\D_{K_2})&
\Sigma_1=\{e,\kappa_3(0)\},\ \Sigma_2=\{e,\kappa_2(1)\},&
\alpha_1=\theta_1,\ \beta_1=\pi-\alpha_1\\
K_1\hbox{ even, }K_2\hbox{ odd}&
\Delta_1=\{e,\kappa_2(0),\kappa_3(0),\kappa_4(-K_1/2,(K_2-1)/2)\},&\\
&\Delta_2=\{e,\kappa_2(1),\kappa_3(0),\kappa_4(K_1/2,(K_2-1)/2)\},&
\alpha_2=\theta_2,\ \beta_2=\pi-\alpha_2\\
&\gamma=((\cos\theta_1,0,0,\sin\theta_1);(\cos\theta_2,0,0,\sin\theta_2)&\\
\hline
\end{array}
$$
\end{sidewaystable}

\begin{sidewaystable}
$$
\begin{array}{|l|l|l|}
\hline
{\Gamma} & \Sigma_j,\ \Delta_j \hbox{ and }\gamma & \alpha_j\hbox{ and }\beta_j\\
\hline
(\D_{2K_1}/\D_{K_1};\D_{2K_2}/\Z_{4K_2})&
\Sigma_1=\{e,\kappa_3(0,0)\},\ \Sigma_2=\{e,\kappa_3(K_1/2,K_2)\}&
\alpha_1=\pi/2,\ \beta_1=\pi/2\\
K_1\hbox{ even}&\Delta_1=\{e,\kappa_1(-),\kappa_3(0,0),\kappa_3(K_1/2,3K_2)\},&\\
&\Delta_2=\{e,\kappa_1(+),\kappa_3(0,0),\kappa_3(K_1/21,K_2)\},\
\gamma=((1,0,0,0);(0,0,0,1))&\alpha_2=\pi/2,\ \beta_2=\pi/2\\
\hline
(\D_{2K_1}/\D_{K_1};\D_{2K_2}/\Z_{4K_2})&
\Sigma_1=\{e,\kappa_1(0)\},\ \Sigma_2=\{e,\kappa_2(1)\}&
\alpha_1=\theta_1,\ \beta_1=\pi-\alpha_1\\
K_1\hbox{ odd}&\Sigma_3=\{e,\kappa_1(1)\},\ \Sigma_4=\{e,\kappa_2(0)\},&
\alpha_2=\theta_2/2,\ \beta_2=\pi-\alpha_2\\
&\Delta_1=\{e,\kappa_1(0),\kappa_2(0),\kappa_3((K_1+1)/2,K_2)\},&\\
&\Delta_2=\{e,\kappa_1(0),\kappa_2(1),\kappa_4((K_1+1)/2,K_2+1)\},&
\alpha_3=\theta_1,\ \beta_3=\pi-\alpha_3\\
&\Delta_3=\{e,\kappa_1(1),\kappa_2(1),\kappa_3((K_1+1)/2+1,K_2+1)\},&\\
&\Delta_4=\{e,\kappa_1(1),\kappa_2(0),\kappa_4((K_1+1)/2+1,K_2)\};\ \gamma=e&
\alpha_4=\theta_2/2,\ \beta_4=\pi-\alpha_4\\
\hline
(\D_{2K}/\D_{2K};\T/\T)&\Sigma_1=\{e,\kappa_2(0,0)\},&
\alpha_1=\pi/4,\ \beta_1=\pi/4\\
K\hbox{ even}&\Delta_1=\{e,\kappa_1(+,2),\kappa_2(0,0),\kappa_2(K,1)\}&\\
&\gamma=((1,0,0,1)/\sqrt{2};(1/2,1/2,1/2,1/2))&\\
\hline
(\D_{2K}/\D_{2K};\T/\T)&\Sigma_1=\{e,\kappa_2(0,0)\},\
\Sigma_2=\{e,\kappa_2(K,1)\}&\alpha_1=\pi/4,\ \beta_1=\pi/4\\
K\hbox{ odd}&\Delta_1=\{e,\kappa_1(+,2),\kappa_2(0,0),\kappa_2(-K,1)\}&
\alpha_2=\pi/4,\ \beta_2=\pi/4\\
&\Delta_2=\{e,\kappa_1(+,2),\kappa_2(0,0),\kappa_2(K,1)\}&\\
&\gamma=((0,0,0,1);(-1/2,1/2,1/2,1/2))&\\
\hline
(\D_{2K}/\D_{2K};\mO/\mO)&\Sigma_1=\{e,\kappa_2(+,0,+)\},\
\Sigma_2=\{e,\kappa_4(0,0,-)\},\ \Sigma_3=\{e,\kappa_4(K,0,+)\},
&\alpha_1=\theta/2,\ \beta_1=\pi-\alpha_1\\
K\hbox{ odd}&\Delta_1=\{e,\kappa_2(+,0,+),\kappa_4(1,0,-),\kappa_3(K-1,0)\},&
\alpha_2=\pi/4,\ \beta_2=3\pi/4\\
&\Delta_2=\{e,\kappa_2(+,0,+),\kappa_4(0,0,-),\kappa_3(K,0)\},&
\alpha_3=\pi/4,\ \beta_3=3\pi/4\\
&\Delta_3=\{e,\kappa_1(+,0),\kappa_4(0,0,-),\kappa_4(K,0,+)\}&\\
&\gamma=((\cos(\theta(K-1)/2),0,0,\sin(\theta(K-1)/2);(0,0,1,0))&\\
\hline
(\D_{2K}/\Z_{4K};\mO/\T)&\Sigma_1=\{e,\kappa_2(0,0,+)\},&\alpha_1=\pi/2,\ \beta_1=\pi/2\\
K\hbox{ even}&\Delta_1=\{e,\kappa_1(-,0),\kappa_2(0,0,+),\kappa_2(-K,0,-)\},&\\
&\gamma=((1,0,0,1)/\sqrt{2};(0,0,0,1))&\\
\hline
(\D_{2K}/\Z_{4K};\mO/\T)&\Sigma_1=\{e,\kappa_2(0,0,+)\},\
\Sigma_2=\{e,\kappa_2(K,0,-)\}&\alpha_1=\pi/2,\ \beta_1=\pi/2\\
K\hbox{ odd}&\Delta_1=\{e,\kappa_1(-,0),\kappa_2(0,0,+),\kappa_2(-K,0,-)\},&\alpha_2=\pi/2,\ \beta_2=\pi/2\\
&\Delta_2=\{e,\kappa_1(+,0),\kappa_2(0,0,+),\kappa_2(K,0,-)\}&\\
&\gamma=((0,0,0,1);(1,0,0,0))&\\
\hline
\end{array}
$$
Continuation of Annex \ref{Sigmaj-Deltaj}
\end{sidewaystable}

\begin{sidewaystable}
$$
\begin{array}{|l|l|l|}
\hline
{\Gamma} & \Sigma_j,\ \Delta_j \hbox{ and }\gamma & \alpha_j\hbox{ and }\beta_j\\
\hline
(\D_{2K}/\D_K;\mO/\T)&
\Sigma_1=\{e,\kappa_3(0,0,+)\}&\alpha_1=\pi/2,\ \beta_1=\pi/2\\
K=4k&\Delta_1=\{e,\kappa_1(-,0),\kappa_3(0,0,+),\kappa_3(-K/2,0,-)\}&\\
&\gamma=((1,0,0,1)/\sqrt{2};(0,1,0,0))&\\
\hline
(\D_{2K}/\D_K;\mO/\T)&
\Sigma_1=\{e,\kappa_2(0,0)\},\
\Sigma_2=\{e,\kappa_3(-(K-1)/2,0,+)\}&\alpha_1=\pi/4,\ \beta_1=\pi/4\\
K\hbox{ odd}&\Delta_1=\{e,\kappa_1(-,0,-),\kappa_2(0,0),\kappa_3(-(K-1)/2,0,+)\},&\alpha_2=\pi/2,\ \beta_2=\pi/2\\
&\Delta_2=\{e,\kappa_1(+,0,-),\kappa_2(0,0),\kappa_3(K,0,-)\}&\\
&\gamma=((0,0,0,1);(1,0,0,-1)/\sqrt{2})&\\
\hline
(\D_{6K}/\Z_{4K};\mO/\V)&
\Sigma_1=\{e,\kappa_2(0,+)\}&\alpha_1=\pi/2,\ \beta_1=\pi/2\\
K\hbox{ even}&\Delta_1=\{e,\kappa_1(+,0),\kappa_2(0,+),\kappa_2(-K,-)\},\
\gamma=((1,0,0,1)/\sqrt{2};(0,1,0,0))&\\
\hline
(\D_{6K}/\Z_{4K};\mO/\V)&
\Sigma_1=\{e,\kappa_2(0,+)\},\
\Sigma_2=\{e,\kappa_2(K,-)\}&\alpha_1=\pi/2,\ \beta_1=\pi/2\\
K\hbox{ odd}&\Delta_1=\{e,\kappa_1(-,0),\kappa_2(0,+),\kappa_2(-K,-)\},&
\alpha_2=\pi/2,\ \beta_2=\pi/2\\
&\Delta_2=\{e,\kappa_1(+,0),\kappa_2(0,+),\kappa_2(K,-)\},
\gamma=((0,0,0,1);(1,0,0,0))&\\
\hline
(\D_{2K}/\D_{2K};\I/\I)&\hbox{ the same as }(\D_{2K}/\D_{2K};\T/\T)&\\
\hline
(\T/\Z_2;\T/\Z_2)&\Sigma_1=\{e,\kappa_1(-,0)\},&\alpha_1=\pi/2,\ \beta_1=\pi\\
&\Delta_1=\{e,\kappa_1(-,0),\kappa_1(-,1),\kappa_1(+,2)\}&\\
&\gamma=((1/2,1/2,1/2,1/2);(1/2,1/2,1/2,1/2))&\\
\hline
(\T/\T;\mO/\mO)&\Sigma_1=\{e,\kappa_2(+,0,0,+)\},&\alpha_1=\pi/4,\ \beta_1=\pi/4\\
&\Delta_1=\{e,\kappa_1(+,1,0),\kappa_2(+,0,0,+),\kappa_2(+,2,0,-)\}&\\
&\gamma=((1/2,1/2,1/2,1/2);(0,0,0,1))&\\
\hline
(\D_{2rK_1}/\Z_{K_1};\D_{2rK_2}/\Z_{K_2})&
\Sigma_1=\{e,\kappa_2(0,0,0)\},\ \Sigma_2=\{e,\kappa_3((K_1-1)/2,(K_2-1)/2,0)\}
&\alpha_1=\pi,\ \beta_1=\pi\\
&\Delta_1=\{e,\kappa_1,\kappa_2(0,0,0),\kappa_3((K_1-1)/2,(K_2-1)/2,0)\},\
\Delta_2=\Delta_1&\alpha_2=\pi,\ \beta_2=\pi\\
&\gamma=e&\\
\hline
\end{array}
$$
Continuation of Annex \ref{Sigmaj-Deltaj}
\end{sidewaystable}

\begin{thebibliography}{99}

\bibitem{am02}
P.~Ashwin and J.~Montaldi.
Group theoretic conditions for existence of robust relative homoclinic trajectories.
{\em Math. Proc. Camb. Phil. Soc.} {\bf 133}, 125 -- 141 (2002).

\bibitem{cl2000}
P.~Chossat and R.~Lauterbach.
Methods in Equivariant Bifurcations and Dynamical Systems. {\em World Scientific Publishing Company}, 2000.

\bibitem{conw}
J.~H.~Conway, D.~Smith. {\em On Quaternions and Octonions.} A K Peters:
Natick, Massachusets, 2003.

\bibitem{pdv}
P.~Du~Val. {\em Homographies, Quaternions and Rotations.} OUP: Oxford, 1964.

\bibitem{fach}
G.~Faye, P.~Chossat.  Bifurcation diagrams and heteroclinic networks of octagonal H-planforms, {\em J. of Nonlinear Science}, {\bf 22}, 1, 277-326 (2012).

\bibitem{Field}
M.~Field. {\em Lectures on Bifurcations, Dynamics, and Symmetry}. Pitman Research Notes in Math. Series 356, Longman,1996

\bibitem{GSS} M.~Golubitsky, I.~Stewart and D.~Schaeffer. {\em {Singularities and Groups in Bifurcation Theory}}.  Springer, 1988.

\bibitem{koenig} M.~Koenig.
Linearization of vector fields on the orbit space of the action of a compact Lie group.
{\em Proc. Camb. Phil. Soc.} {\bf 121}, 401--424 (1997).

\bibitem{Kru97}
M.~Krupa. Robust heteroclinic cycles.
{\em J. Nonlinear Science}, {\bf 7}, 129 -- 176 (1997).

\bibitem{km95a}
M.~Krupa and I.~Melbourne.
Asymptotic stability of heteroclinic cycles in systems with symmetry.
{\em Ergodic Theory Dyn. Syst.} {\bf 15}, 121 -- 148 (1995).

\bibitem{km04}
M.~Krupa and I.~Melbourne.
Asymptotic stability of heteroclinic cycles in systems with symmetry.~II.
{\em Proc. Roy. Soc. Edinburgh} {\bf 134A}, 1177 -- 1197 (2004).

\bibitem{Lang}
S.~Lang. {\em Algebra, third edition.} Addison-Wesley, 1993.

\bibitem{laumat} R.~Lauterbach and P.~Matthews. Do absolutely irreducible group actions
have odd dimensional fixed point spaces? arXiv:1011.3986v1 (2010).

\bibitem{melb} I.~Melbourne. An example of a nonasymptotically stable attractor. {\em Nonlinearity} {\bf 4}, 835  -- 844(1991).

\bibitem{op12}
O.M.~Podvigina. Stability and bifurcations of heteroclinic cycles of type Z.
{\it Nonlinearity} {\bf 25}, 1887 -- 1917, arXiv:1108.4204 [nlin.CD] (2012).

\bibitem{op13}
O.M.~Podvigina. Classification and stability of simple homoclinic cycles in
$\R^5$. {\it Nonlinearity} {\bf 26}, 1501 -- 1528, arXiv:1207.6609 [nlin.CD] (2013).

\bibitem{pa11}
O.M.~Podvigina and P.~Ashwin.
On local attraction properties and a stability index for heteroclinic
connections. {\it Nonlinearity} {\bf 24}, 887 -- 929, arXiv:1008.3063 [nlin.CD] (2011).


\bibitem{schwarz}
G.~W.~Schwarz. Lifting smooth homotopies of orbit spaces.
{\em Publ. Math. I.H.E.S.} {\bf 51}, 37 -- 135 (1980).

\bibitem{sot03}
N.~Sottocornola. Robust homoclinic cycles in $\R^4$.
{\it Nonlinearity} {\bf 16}, 1 -- 24 (2003).

\bibitem{sot05}
N.~Sottocornola. Simple homoclinic cycles in low-dimensional spaces.
{\it J.~Differential Equations} {\bf 210}, 135 -- 154 (2005).

\end{thebibliography}
\end{document}